\documentclass[seceq,preprint]{ptptex}
\usepackage{graphicx}
\usepackage{wrapft}

\preprintnumber[3cm]{KEK-CP-162}

\title{
 Effect of low-lying fermion modes in the $\epsilon$-regime
  of QCD 
}

\newcommand{\GUAS}{
  School of High Energy Accelerator Science,
  The Graduate University for Advanced Studies (Sokendai),
  Tsukuba, Ibaraki 305-0801, Japan
}
\newcommand{\KEK}{
  High Energy Accelerator Research Organization (KEK),
  Tsukuba, Ibaraki 305-0801, Japan
}

\author{
Kenji \textsc{Ogawa}$^1$%
and
Shoji \textsc{Hashimoto}$^{1,2}$%
}
\inst{
$^1$\GUAS
\\
$^2$\KEK
}

\abst{
  We investigate the effects of low-lying fermion eigenmodes  
  on the QCD partition function in the $\epsilon$-regime. 
  The fermion determinant is approximated by a truncated
  product of low-lying eigenvalues of the overlap-Dirac
  operator.
  With two flavors of dynamical quarks, we observe that the
  lattice results for the lowest eigenvalue distribution,
  eigenvalue sum rules and partition function reproduce the
  analytic predictions made by Leutwyler and Smilga, which
  strongly depend on the topological charge of the
  background gauge configuration. 
  The value of chiral condensate extracted from these
  measurements are consistent with each other.
  For one dynamical quark flavor, on the other hand, we find
  an apparent disagreement among different determinations of
  the chiral condensate, which may suggest the failure of the
  $\epsilon$-expansion in the absence of massless
  Nambu-Goldstone boson.
}

\begin{document}

\date{\today}
\maketitle

\section{Introduction}
Chiral perturbation theory (ChPT) is an effective field
theory to describe the dynamics of Quantum Chromodynamics
(QCD) at low energies ($\ll\Lambda\sim$ 1~GeV), where the
Nambu-Goldstone pion excitations dominate the dynamics while
other particles are too heavy to be excited.
In the infinite volume, ChPT provides a method to express
low energy amplitudes of pion as an expansion in terms of
pion mass squared $m_\pi^2$ and its momentum squared $p^2$.
Thus it enables us to calculate low energy pion
amplitudes in a systematic manner. 
When the system is put in a finite volume,
for which the pion Compton wavelength ($\sim 2\pi/m_\pi$) is
much larger than the linear extent $L$ of the space-time, 
{\it i.e.} $m_\pi L\ll$~1, 
while $L$ is kept large enough compared to the QCD scale
$1/\Lambda_{\mathrm{QCD}}$,
the low energy effective theory can still be constructed as
an expansion in terms of small 
$\epsilon^2\sim m_\pi/\Lambda\sim p^2/\Lambda^2$,
which is known as the $\epsilon$-expansion
\cite{Gasser:1987ah}.
In this setup, the so-called $\epsilon$-regime of ChPT, the chiral
symmetry is not spontaneously broken and one must explicitly 
integrate over different vacua in the path integral, which
leads to characteristic behavior of the partition function
and other physical quantities.
In particular, they strongly depend on the gauge field
topology (or the existence of fermion zero modes)
\cite{Leutwyler:1992yt}.

Lattice QCD simulation is well suited to the study of finite 
volume physics.
Since the chiral symmetry plays an essential role in the
$\epsilon$-regime, one should use lattice fermion
formulations that respect chiral symmetry.
Such fermion formulation has become available relatively
recently, 
{\it i.e.} the overlap fermion
\cite{Neuberger:1997fp,Neuberger:1998wv} and 
the domain-wall fermion \cite{Kaplan:1992bt,Shamir:1993zy}.
These fermion formulations satisfy the Ginsparg-Wilson
relation \cite{Ginsparg:1981bj}, with which the fermion
action can be shown to have exact chiral symmetry at finite
lattice spacings \cite{Luscher:1998pq}.
Quenched lattice simulations have so far been done in the
$\epsilon$-regime to calculate the chiral condensate
\cite{Hernandez:1999cu} and other low energy constants
\cite{Bietenholz:2003bj,Giusti:2003iq,Giusti:2004yp,Fukaya:2005yg}.
It has also been shown that the distribution of low-lying
eigenvalues of the lattice overlap Dirac operator agrees
well with expectations from the chiral Random Matrix Theory
(RMT)  
\cite{Edwards:1999ra,Bietenholz:2003mi,Giusti:2003gf}
(for a review of RMT, see \cite{Verbaarschot:2000dy}),
which is equivalent to the chiral Lagrangian at the leading
order of the $\epsilon$-expansion.

In this work we extend these previous lattice studies to the
case of non-zero dynamical fermion flavors.
In particular we calculate the topological susceptibility
and partition function with one and two flavors of dynamical
quarks, and investigate their dependence on the quark mass.
The predictions from ChPT in the $\epsilon$-regime are
available from the work by Leutwyler and Smilga
\cite{Leutwyler:1992yt}. 
Due to the presence of fermion zero-modes the partition
function is drastically different for different topological
sectors of background gauge field.
We therefore calculate them for each topological sector
identified by the number of fermion zero-modes.
We also investigate the sum rules of Dirac operator
eigenvalues (the so-called Leutwyler-Smilga sum rules
\cite{Leutwyler:1992yt}) derived from the quark mass
dependence of the partition function.
For these quantities we found that the lattice data with one
or two fermion flavors are well described by the known
analytic formulas. 

We also extend the comparison of low-lying eigenvalue
distributions with the RMT results to the case of non-zero
number of flavors. 
In RMT there is an universality among the number of flavor
$N_f$ and topological charge $\nu$.
Namely the prediction depends only on the combination
$N_f+|\nu|$, which should be observed in the lattice data.

In order to investigate the effect of dynamical fermions, we
employ an approximation for the fermion determinant.
Namely, we approximate the fermion determinant by a product
of low-lying eigenvalues of the overlap Dirac operator.
The effects of higher fermion modes are neglected, which we
call the truncated determinant approximation.
Since the eigenvalues much higher than the quark mass and
the QCD scale $\Lambda_{\mathrm{QCD}}$ should be irrelevant
to low energy physics, this approximation is expected to be
effective for the study of the $\epsilon$-regime.
The higher eigenmodes are sensitive to lattice artifacts and
their main effect for low energy physics appears through the
renormalization of parameters in the theory, {\it i.e.} in
this case the gauge coupling constant and quark masses. 
As we neglect such effects, we effectively choose a different
renormalization scheme.
To what extent this approximation works can be tested by
varying the cutoff for the Dirac operator eigenvalues.

This paper is organized as follows.
First, in Section~\ref{sec:Leutwyler-Smilga} we briefly
review the analytic expectations from ChPT following the
seminal paper by Leutwyler and Smilga
\cite{Leutwyler:1992yt}. 
We then explain our calculation methods and describe the
truncated determinant approximation in some detail
in Sections~\ref{sec:Lattice_details} and
\ref{sec:Truncated_determinant}, respectively.
Numerical results are presented in
Section~\ref{sec:Numerical_results}. 
An earlier presentation of this work is found in
\cite{Ogawa:2004ru}.

\section{Leutwyler-Smilga's analytic predictions}
\label{sec:Leutwyler-Smilga}
In this section we briefly review the analytic predictions
made by Leutwyler and Smilga \cite{Leutwyler:1992yt} using
the chiral Lagrangian in the $\epsilon$-regime.

At the leading order in the $\epsilon$-expansion the kinetic
term in the chiral Lagrangian is suppressed, because the
momentum excitation energy is large in a small volume even
for the unit momentum $2\pi/L$.
The partition function is then given by 
\begin{equation}
  \label{eq:partition_function_Nf=1}
  Z = \exp \left[\Sigma V \mathrm{Re}(me^{i\theta})\right]
\end{equation}
for $N_f$ = 1 and 
\begin{equation}
  \label{eq:partition_function_Nf}
  Z = \int_{\mathrm{SU}(N_f)} d\mu(U_0)
  \exp\left[ \Sigma V 
    \mathrm{Re}[me^{i\theta/N_f} \mathrm{Tr}(U_0^\dagger)]
  \right]
\end{equation}
for $N_f\ge 2$.
Here, $\Sigma$ represents the chiral condensate and $V$ is the 
space-time volume.
They appear in the combination $m\Sigma V$ together with the
quark mass $m$.
The integral in (\ref{eq:partition_function_Nf})
runs over compact $\mathrm{SU}(N_f)$ elements $U_0$
corresponding to the zero momentum mode wave function,
and $d\mu(U_0)$ is its Haar measure.
This integral is necessary because the chiral symmetry is not
spontaneously broken in the finite volume. 
For $N_f$ = 1 (Eq.~(\ref{eq:partition_function_Nf=1})), on
the other hand, this degree of freedom does not remain due to
the U(1) chiral anomaly. 
The parameter $\theta$ is the CP angle of the QCD vacuum.

The partition function in each topological sector labeled by 
the topological charge $\nu$ is obtained by Fourier
transforming (\ref{eq:partition_function_Nf=1}) and
(\ref{eq:partition_function_Nf}).
They are
\begin{align}
  \label{eq:Z_nu_Nf=1}
  Z_\nu & = I_\nu(x) 
  & \mbox{for $N_f$ = 1},
  \\
  \label{eq:Z_nu_Nf=2}
  Z_\nu & = I_\nu^2(x) - I_{\nu+1}(x) I_{\nu-1}(x)
  & \mbox{for $N_f$ = 2}.
\end{align}
$I_\nu(x)$ is the modified Bessel function and its argument
$x$ is $x\equiv m\Sigma V$.
Near the massless limit it behaves as
$(x/2)^{|\nu|}/|\nu!|$.

The probability to find a configuration with a given
topological charge $\nu$ is given by $Z_\nu/Z$, where
$Z=\sum_\nu Z_\nu$ is 
$e^x$ and $I_1(2x)/x$ for $N_f$ = 1 and 2,
respectively.
The topological susceptibility $\langle\nu^2\rangle/V$ is
then obtained through the definition
$\langle\nu^2\rangle = \sum_\nu \nu^2 Z_\nu/Z$ as
\begin{align}
  \label{eq:tpsus_nf1}
  \frac{\langle\nu^2\rangle}{V} & = m \Sigma
  & \mbox{for $N_f$ = 1},
  \\
  \label{eq:tpsus_nf2}
  \frac{\langle\nu^2\rangle}{V} &
  = \frac{m\Sigma I_2(m\Sigma V)}{2 I_1(m\Sigma V)}
  & \mbox{for $N_f$ = 2}.
\end{align}
Asymptotically, it behaves as 
$\langle\nu^2\rangle/V \varpropto m^{N_f}$ for $x\ll 1$ 
and
$\langle\nu^2\rangle/V=m\Sigma/N_f$ for $x\gg 1$.

The above results for the partition function should agree
with the partition function of the underlying theory, QCD,
in the appropriate limit.
The QCD partition function for a given topological charge
$\nu$ is written as
\begin{equation}
  \label{eq:Znu_QCD}
  Z_\nu = m^{N_f|\nu|} \int_\nu [dG] e^{-S_G}
  \prod'_n (|\lambda_n|^2+m^2)^{N_f}.
\end{equation}
The path integral over the gauge field configuration is done
for a fixed topology $\nu$ with a weight given by the gauge 
action $S_G$.
The fermion determinant is written in the form of a product
of the eigenvalues $\lambda_n$ of the Dirac operator.
Non-zero eigenvalues appear with their complex conjugate and
the index $n$ in (\ref{eq:Znu_QCD}) runs over eigenvalues
with positive imaginary part only.
The prime on the product indicates that it excludes
the zero modes, which are factored out as $m^{N_f|\nu|}$.
Defining the expectation value in the massless limit as
$\langle\!\langle\cdots\rangle\!\rangle_\nu$, the above
expression is rewritten as
\begin{equation}
  \frac{m^{-N_f|\nu|}Z_\nu(m)}{
    \lim_{m\rightarrow 0}\, m^{-N_f|\nu|}Z_\nu(m) } 
  =
  \left\langle\!\!\!\left\langle \prod'_n 
  \left(1+\frac{m^2}{|\lambda_n|^2}\right)^{N_f}
  \right\rangle\!\!\!\right\rangle_\nu.
\end{equation}
In the effective theory the same quantity is given by
\begin{align}
  |\nu|! \left(\frac{2}{x}\right)^{|\nu|} I_\nu(x)
  & &\mbox{for $N_f$ = 1},
  \\
\left [ \left( \frac{1}{\nu !} \right ) ^2 - \frac{1}{(\nu + 1)! |\nu-1|!} \right ]^{-1} 
\left(\frac{2}{x} \right)^{2|\nu|}
Z_\nu(x)
  & &\mbox{for $N_f$ = 2}.
\end{align}
By taking a derivative with respect to $m^2$ for both QCD
and the effective theory and equating them at $m=0$, we arrive
at the so-called Leutwyler-Smilga sum rules.
From the first derivative we obtain
\begin{equation}
  \label{eq:LS_sum-rule_1}
  \left\langle\!\!\!\left\langle
      \sum_n' \frac{1}{(|\lambda_n|\Sigma V)^2}
    \right\rangle\!\!\!\right\rangle_\nu 
  =
  \frac{1}{4(N_f+|\nu|)},
\end{equation}
which we call the sum rule I in this paper.
The prime on the sum implies that it does not include the
zero modes.
For non-degenerate flavors with masses $m_i$ and $m_j$ we
can extract two sum rules from second derivatives
corresponding to $m_i^2 m_j^2$ and $m_i^4$
\begin{eqnarray}
  \label{eq:LS_sum-rule_2a}
  \left\langle\!\!\!\left\langle
      \left[\sum_n' \frac{1}{(|\lambda_n|\Sigma V)^2}\right]^2
    \right\rangle\!\!\!\right\rangle_\nu            
  & = &
  \frac{1}{16((N_f+|\nu|)^2-1)},
  \\
  \label{eq:LS_sum-rule_2b}
  \left\langle\!\!\!\left\langle
      \left[\sum_n' \frac{1}{(|\lambda_n|\Sigma V)^2}\right]^2
      - \sum_n' \frac{1}{(|\lambda_n|\Sigma V)^4}
    \right\rangle\!\!\!\right\rangle_\nu            
  & = &
  \frac{1}{16(N_f+|\nu|+1)(N_f+|\nu|+2)}
\end{eqnarray}
which we call the sum rules IIa and IIb, respectively.
Their difference gives
\begin{equation}
  \label{eq:LS_sum-rule_2c}  
  \left\langle\!\!\!\left\langle
      \sum_n' \frac{1}{(|\lambda_n|\Sigma V)^4}
    \right\rangle\!\!\!\right\rangle_\nu 
   = 
  \frac{1}{16(N_f+|\nu|)((N_f+|\nu|)^2-1)},
\end{equation}
(sum rule IIc).
Note that there is a universality among the number of
flavors and topological charge, 
{\it i.e.} they appear only in the combination $N_f+|\nu|$. 

These sum rules
(\ref{eq:LS_sum-rule_1})-(\ref{eq:LS_sum-rule_2c})
do not make sense as they stand, because the left hand side 
is ultraviolet divergent while the right hand side is a
definite number.
In fact, since the eigenvalue density
$\rho(\lambda)=\frac{1}{V}
\left\langle \sum_n\delta(\lambda_n-\lambda)\right\rangle$ 
increases as $\sim V\lambda^3$ for large $\lambda$, the sum
rule I is quadratically divergent.
The sum rules IIa and IIb have quartic divergences, which
cancel in the rule IIc and only a logarithmic divergence
remains. 
In spite of these divergences, the sum rules
(\ref{eq:LS_sum-rule_1})-(\ref{eq:LS_sum-rule_2c}) 
are consistent, because they only affect the corrections of
order $1/V$ or higher, and thus can be removed by first
taking a limit $V\to\infty$.
In the following numerical work we do not perform this
extrapolation, which is very expensive, but consider 
differences between different topological sectors.
Since the eigenvalue density becomes independent of topology
for large $\lambda$, the ultraviolet divergence cancels in
such differences.

\section{Lattice details}
\label{sec:Lattice_details}
The chiral symmetry is essential for the lattice study in
the $\epsilon$-regime.
We use the Neuberger (or the so-called overlap) Dirac
operator \cite{Neuberger:1997fp,Neuberger:1998wv} which
respects the Ginsparg-Wilson relation, and thus the chiral
symmetry is preserved at finite lattice spacing.

The overlap Dirac operator $D$ is defined as
\begin{equation}
  \label{eq:overlap}
  D  = \frac{1}{\bar a} 
  \left[ 1 + \gamma _5 \mathrm{sgn}(a H_W) \right],
\end{equation}
with
\begin{equation}
  H_W = \gamma _5 \left[ {D_W  - \frac{1}{\bar a} } \right],
\end{equation}
\begin{equation}
  D_W  = \frac{1}{2} \sum_\mu 
  \left\{ 
    \gamma_\mu (\nabla_\mu^* + \nabla_\mu)
    - a\nabla_\mu^*\nabla_\mu
  \right\},
\end{equation}
and $\bar a = a/(1+s)$.
Here $D_W$ is the conventional Wilson-Dirac operator and
$\nabla$ ($\nabla^*$) denotes a forward (backward)
gauge-covariant differential operator on the lattice.
The parameter $s$ is introduced to optimize the negative
mass term given in the kernel $H_W$.
In this work we choose $s$ = 0.6 at $\beta$ = 5.85.
The sign function in (\ref{eq:overlap}) is approximated
using the Chebyshev polynomial after subtracting out several 
lowest eigenmodes of $H_W$.
The order of polynomial is optimized for each gauge
configuration to ensure the accuracy of $10^{-11}$ 
for the sign function, and it is typically around 100--200
when we subtract 14 lowest-lying eigenvalues of $H_W$.

We work on quenched gauge configurations generated with the
plaquette gauge action at $\beta$ = 5.85 on a $10^4$
lattice. 
The lattice spacing determined through the Sommer scale
$r_0$ is 0.123~fm.
The topological charge for these gauge configurations is
determined by the number of zero modes of the overlap-Dirac
operator (\ref{eq:overlap}).
The number of gauge configuration for each topological
sector is given in Table~\ref{tab:config}.

\begin{table}[tbp]
\begin{center}
  \begin{tabular}{cccccccc}\hline \hline
    Topological charge $|\nu|$ & 
    \makebox[5ex]{0} & \makebox[5ex]{1} & \makebox[5ex]{2} & 
    \makebox[5ex]{3} & \makebox[5ex]{4} & \makebox[5ex]{5} & 
    \makebox[5ex]{6} 
    \\ 
    \#config & 169 & 290  & 149 & 68 & 20 & 4 & 3

\\ \hline \hline
  \end{tabular}
  \caption{The number of quenched gauge configurations for
    each topological sector.}
  \label{tab:config}
\end{center}
\end{table}

For the overlap-Dirac operator (\ref{eq:overlap}) we
calculate 50 lowest eigenvalues and their eigenvectors for
each gauge configuration.
We utilize the numerical package ARPACK \cite{ARPACK}, which
implements the implicitly restarted Arnordi method to
compute the eigenvalues.
Instead of treating the operator $D$ as it is, we consider a
chirally projected operator $D^\pm\equiv P_\pm DP_\pm$, where 
$P_\pm\equiv (1\pm\gamma_5)/2$.
The size of the matrix is then a factor of 2 smaller and the
numerical calculation becomes about 2 times faster.
The eigenvalues of $D$ can be obtained from the eigenvalues
of $P_\pm DP_\pm$, using the property that the chirally
projected operator gives a real part of the eigenvalues of
the original operator. 
Since the eigenvalues of $D$ lie on a circle satisfying
$|1-\lambda|^2=1$, we can construct a pair of eigenvalues of
$D$, namely $\lambda$ and $\lambda^*$.
The corresponding eigenvectors $u$ of $D$ can also be
reconstructed from the eigenvectors $u^\pm=P_\pm u$ of the
chirally projected operator $D^\pm$ using a formula
\begin{equation}
  u = \frac{D-\lambda^*}{\mathrm{Im}\lambda} u^+
    = \frac{D-\lambda^*}{\mathrm{Im}\lambda} u^-.
\end{equation}
In order to identify the number of left-handed and
right-handed zero modes, we have to compute the eigenvalues
of both $P_+DP_+$ and $P_-DP_-$.
We carry out such calculation until we achieve enough
precision to identify pairing non-zero eigenvalues of
both operators \cite{Giusti:2002sm}.
The number of zero modes is then determined unambiguously,
and we continue the calculation of the rest of the
eigenvalues on the operator for which we do not find the
zero modes.

\section{Truncated determinant approximation}
\label{sec:Truncated_determinant}
In our calculation the effects of dynamical fermions are
incorporated in the partition function by including a
product of eigenvalues of Dirac operator in the pure gauge
path integral.
\begin{equation}
  \label{eq:Z_nu_lattice}
  Z_\nu 
  = m^{N_f |\nu|} \int [dU]_\nu e^{-S_G} \prod_n ({\bar \lambda_n}^2+m^2)^{N_f} 
  = m^{N_f |\nu|} \left\langle \prod_n ({\bar \lambda_n}^2+m^2)^{N_f} \right\rangle^Q_\nu
\end{equation}
where $\bar \lambda_n = |\lambda_n| \sqrt{1-(\bar{a}m)^2/4}$.
The eigenvalues of the overlap-Dirac operator appear in
complex conjugate pairs $(\lambda,\lambda^*)$ except for the
zero-modes as in the continuum case.
The notation $\langle\cdots\rangle^Q_\nu$ denotes an
expectation value evaluated on quenched gauge configurations
with a fixed topological charge $\nu$.
Namely, we generate the gauge configurations with a weight
determined by the pure gauge action and reweight them with
the product of eigenvalues.

This method requires a calculation of all eigenvalues of the
overlap-Dirac operator for each configuration, but it is not
feasible unless the lattice size is extremely small 
($\sim 4^4$).
Instead, we approximate the fermion determinant by a
truncated product of eigenvalues in (\ref{eq:Z_nu_lattice}).
The upper limit of the eigenvalue plays a role of an
ultraviolet cutoff for the fermionic degrees of freedom.
Qualitatively, this procedure is justified, since the
physical effects of the fermion determinant are expected to
come from the low-lying eigenmodes, while the higher
eigenmodes should be irrelevant for low energy physics.
As in the usual renormalization program, cutoff dependence
of physical quantity can be absorbed in the coupling
constants and anomalous dimensions 
up to lattice artifacts.
In our case the bulk of the effects of higher fermion modes 
is neglected, and thus the relation between the lattice bare
coupling and the lattice spacing is mostly the same as in
the quenched approximation.

The role of the cutoff in the fermion determinant can be
rephrased by defining a new Dirac operator $D_{\mathrm{LM}}$
(LM stands for low-lying modes) as
\begin{equation}
  \label{eq:D_LM}
  D_{\mathrm{LM}} = \gamma_5 f(H,\lambda_{\mathrm{cutoff}})
\end{equation}
with $H=\gamma_5 D$ and a function
$f(x,\lambda_{\mathrm{cutoff}})$ satisfying a condition
\begin{equation}
  f(x,\lambda_{\rm cutoff})=
  \begin{cases}
    x & (|x| \ll \lambda_{\mathrm{cutoff}}), \\
    \lambda_{\mathrm{cutoff}} & (|x| \gg \lambda_{\mathrm{cutoff}}).
  \end{cases}
\end{equation}
The truncation corresponds to a non-smooth function, which
transfers 
from $x$ to a constant $\lambda_{\mathrm{cutoff}}$ at
$\lambda_{\mathrm{cutoff}}$.
One can also interpolate the two regions smoothly using an
analytic function, {\it e.g.} 
$f(x,\lambda_{\rm cutoff})=\lambda_{\mathrm{cutoff}}
\tanh(x/\lambda_{\mathrm{cutoff}})$.
For such analytic functions the locality and unitarity of
the new Dirac operator $D_{\mathrm{LM}}$ can be proved
\cite{Borici:2002rq}.
Furthermore, the operator $D_{\mathrm{LM}}$ satisfies a
modified Ginsparg-Wilson relation
\begin{equation}
  \gamma_5 D_{\mathrm{LM}} + D_{\mathrm{LM}} \gamma_5 
  = \bar a D \gamma_5 D_{\mathrm{LM}}
  = \bar a D_{\mathrm{LM}} \gamma_5 D,
\end{equation}
as far as $f(x,\lambda_{\rm cutoff})$ is written in terms of
a polynomial.
It implies that the fermion action constructed with the
truncated operator $D_{\mathrm{LM}}$ is invariant under the
chiral transformation
$\delta\psi=\gamma_5(1-\frac{\bar a D}{2})\psi$ and
$\delta\bar{\psi}=\bar{\psi}(1-\frac{\bar a D}{2})\gamma_5$, just
like the original overlap-Dirac operator.
The truncated operator $D_{\mathrm{LM}}$ may, therefore, be
used as an alternative definition of the Dirac operator with
exact chiral symmetry.
It should be noted, however, that our choice for the
function $f(x,\lambda_{\rm cutoff})$ (hard cutoff) has to be
understood as an approximation as it is not a smooth
analytic function.

\begin{figure}[tbp]
  \centering
  \includegraphics[width=10cm]{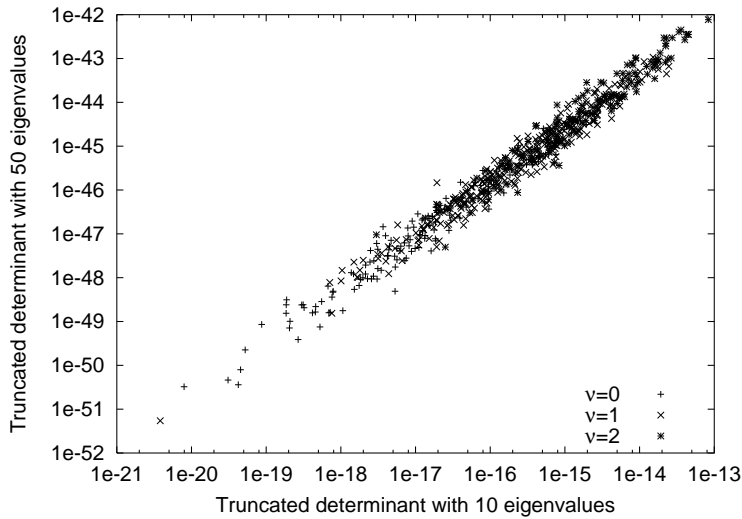}
  \caption{
    Correlation between the truncated determinant
    (\ref{eq:truncated_determinant})
    with $N_{\mathrm{cut}}=10$ (horizontal axis) and that
    with $N_{\mathrm{cut}}=50$ (vertical).
    Each point corresponds to a gauge configuration with
    topological charge 0 (pluses), 1 (crosses), and 2
    (stars).
    The quark mass is set equal to zero.
    Zero-modes for nontrivial topological sectors are not
    included in the product.
  }
  \label{fig:det10_det50}
\end{figure}

On a $10^4$ lattice at $\beta$ = 5.85, we calculated
50 smallest eigenvalues of the chirally projected
overlap-Dirac operator. 
Among them, the maximum eigenvalue corresponds to the
physical scale $\sim$ 1200~MeV, which is large enough as a
cutoff for the study of the low energy physics of interest. 
We define the truncated determinant as
\begin{equation}
  \label{eq:truncated_determinant}
  \overline{\det (D+m)} \equiv
  \prod_{n=1}^{N_{\mathrm{cut}}}({\bar{\lambda}_n}^2+m^2)
\end{equation}
with a fixed number $N_{\mathrm{cut}}$ of eigenvalues.
Figure~\ref{fig:det10_det50} shows a correlation
between the truncated determinant with 
$N_{\mathrm{cut}}$ = 10 and that with 
$N_{\mathrm{cut}}$ = 50 for topological sectors 
$\nu$ = 0, 1, and 2. 
The quark mass is set to zero, and the zero-modes are
factored out for non-trivial topological sectors.
We observe that they are strongly correlated up to an order
of magnitude variations, 
while the truncated determinant itself varies as much as 8
orders of magnitude when measured on quenched gauge
configurations. 
It suggests that the truncated determinant provides a good
approximation of the full determinant already at
$N_{\mathrm{cut}}$ = 10.

\begin{figure}[tbp]
  \centering
  \includegraphics[width=10cm]{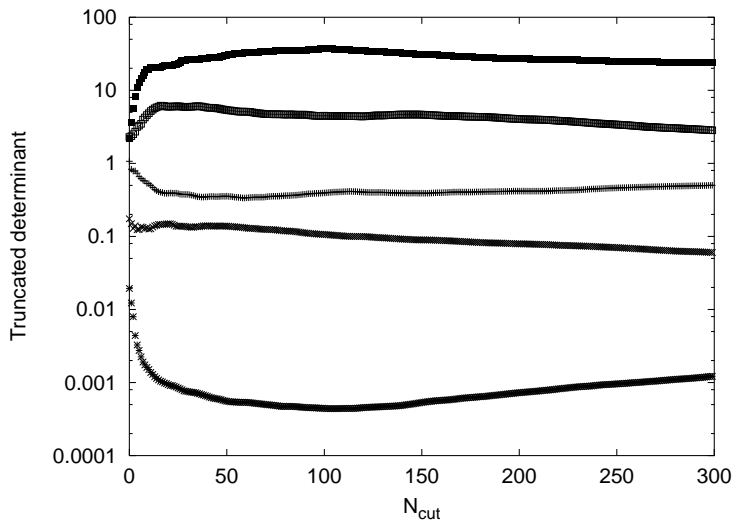}
  \caption{
    Relative weight due to the truncated determinant for the
    massless overlap-Dirac operator as a function of
    $N_{\mathrm{cut}}$ for five representative gauge
    configurations with $\nu=0$. 
  }
  \label{fig:ratio}
\end{figure}

One may also see how much the relative size of the truncated
determinant depends on the value of cutoff
$N_{\mathrm{cut}}$.
In Figure~\ref{fig:ratio}, we plot the truncated determinant
for the massless overlap-Dirac operator 
as a function of $N_{\mathrm{cut}}$
for five representative gauge configurations, for
which we calculated 300 smallest eigenvalues.
The value is normalized by an average over these five
configurations at a given value of $N_{\mathrm{cut}}$, and
thus the plot shows a relative size of 
the weight factor due to the approximate determinant.
We find that the relative weight varies rapidly for small
$N_{\mathrm{cut}}$ ($\sim$ 10), while it becomes roughly
constant above $N_{\mathrm{cut}}\gtrsim$ 50.
It supports our expectation that the truncated determinant
is a good approximation of the full determinant for
$N_{\mathrm{cut}}\simeq$ 50.
In the following studies we also test how much our results
depend on the parameter $N_{\mathrm{cut}}$ for each quantity
we measure.

The number of necessary eigenvalues to be included in the
truncated determinant is expected to be unchanged even at
smaller lattice spacings as long as the physical volume is
kept fixed.
This is because the the eigenvalue density $\rho(\lambda)$
depends on the physical parameters $\Sigma$ and $V$ but not
on the lattice cutoff $1/a$ (up to lattice artifacts).
It depends on the physical volume linearly, and therefore
the number of eigenvalues to be calculated would become
prohibitively large, {\it e.g.} about 14 times larger on a
(2~fm)$^3\times$(4~fm) lattice, which is a typical lattice
size used for hadron spectrum calculations.

The truncated determinant was previously introduced by
Duncan, Eichten and Thacker to develop an efficient
algorithm for light dynamical quarks with the Wilson fermion 
\cite{Duncan:1998gq}.
They also proposed a method to include the higher
fermion modes to make the algorithm exact using the
multiboson technique of L\"uscher \cite{Luscher:1993xx}.
A similar method can also be applied in our case, but the
effective bosonic action becomes non-local with the
overlap-Dirac operator and the simple heat-bath updation
would be impractical.

\begin{figure}[tbp]
\begin{center}
\includegraphics[width=10cm]{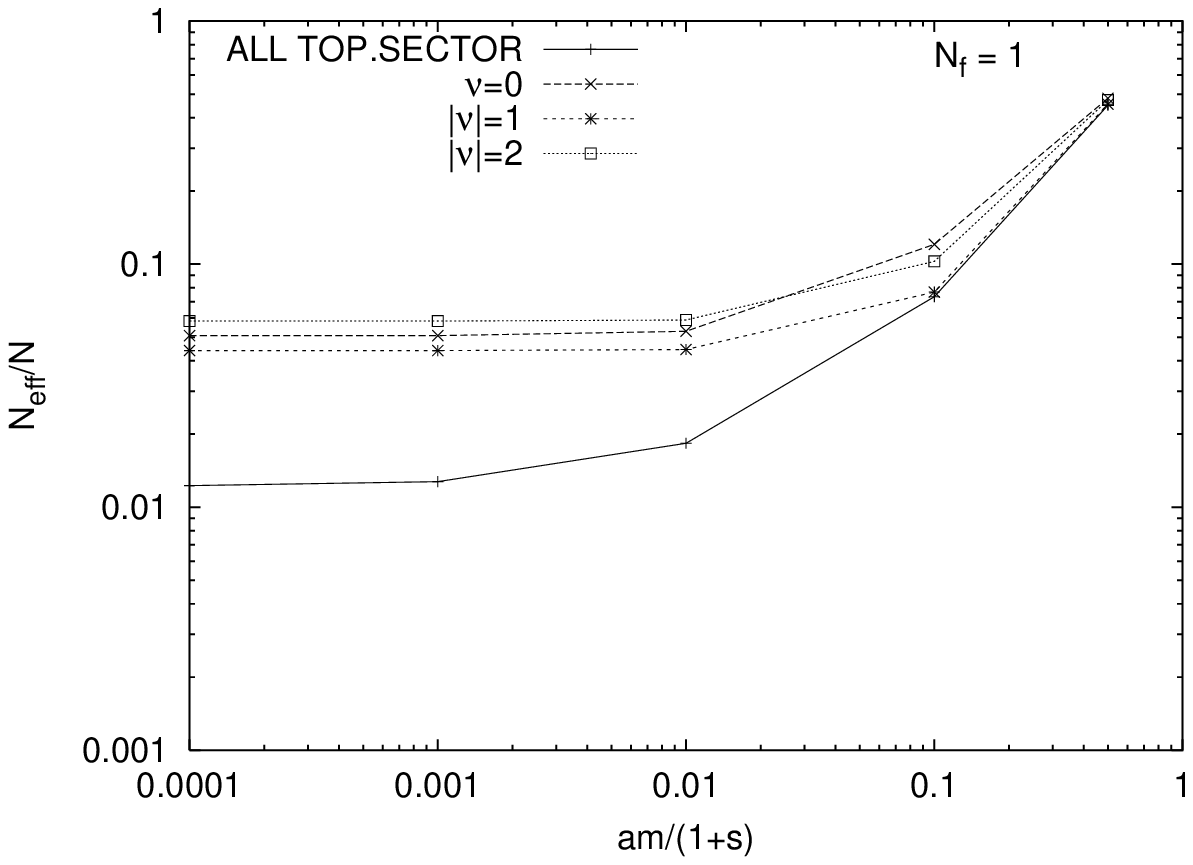}\\
\includegraphics[width=10cm]{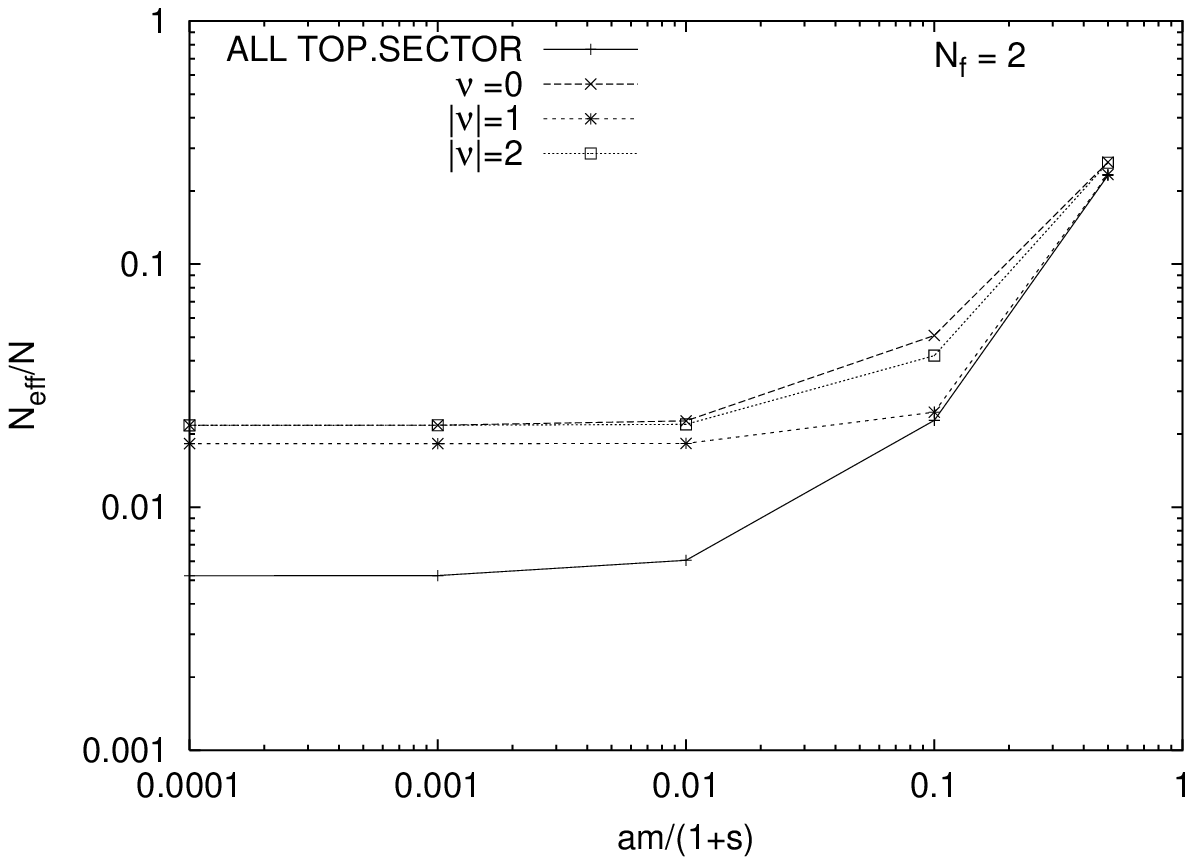}
\caption{
  Effective number of statistics $N_{\rm eff}$ normalized by
  the quenched statistics $N$ for $N_f$ = 1 (top panel) and
  $N_f$ = 2 (bottom panel).
  It is calculated for each topological sector (squares,
  crosses and bursts for $|\nu|$ = 0, 1, and 2,
  respectively) and for all topological sectors (pluses).
}
\label{fig:neff}
\end{center}
\end{figure}

An immediate problem of the reweighting with the approximate 
determinant is that the overlap of the unquenched vacuum
with the quenched vacuum could be tiny and thus the Monte
Carlo sampling becomes inefficient especially for small
quark masses we are interested in.
If this happens, only a few configurations give a dominant
contribution to the Monte Carlo average and others are
suppressed by the reweighting factor representing the
approximate determinant.
We may quantify this effect by considering an effective
number of statistics $N_{\mathrm{eff}}$ as
\begin{equation} 
  N_{\mathrm{eff}} =
  \frac{
    \sum_{i=1}^N \overline{\det D}^{(i)}
  }{
    \max_j (\overline{\det D}^{(j)})
  }.
\end{equation}
Here, $i$ and $j$ label gauge configurations and $N$ is the
number of the configurations. 
$\overline{\det D}^{(i)}$ denotes the truncated determinant
for the $i$-th gauge configuration, and
$\max_j (\overline{\det D}^{(j)})$ is its maximum value in
the gauge ensemble.
Figure \ref{fig:neff} shows the effective number of
statistics (normalized by the quenched statistics) as a
function of quark mass.
The effective number of statistics decreases for smaller
quark masses as expected.
For a given topological sector, it goes down to $\sim$ 0.05
(0.02) for $N_f$ = 1 (2), when the quark mass is decreased
to $\bar{a}m\sim$ 0.01, which corresponds to $m\sim$ 26~MeV.
If we combine all topological charges, we loose another
factor of $\sim$ 5, because only the configurations with
zero topological charge contribute for small quark masses.
Therefore, the number of configurations we are using
(Table~\ref{tab:config}) are not large enough for precision
study of unquenched lattices.

\section{Numerical results}
\label{sec:Numerical_results}
In this section we present our numerical results for the 
eigenvalues of the overlap-Dirac operator.
An eigenvalue $\lambda$ of the overlap-Dirac operator lies
on a circle defined by $|1-a\lambda|=1$.
In the continuum limit $a\to 0$ they become pure imaginary,
corresponding to the eigenvalues of the continuum Dirac
operator.
At the finite lattice spacings, on the other hand, there is
a freedom to define the eigenvalue projected onto the
imaginary axis. 
We simply take an absolute value of the eigenvalue of the
overlap-Dirac operator.
The ambiguity in the choice of the projection leads to a
systematic uncertainty of order $(a\lambda)^2$.
In the following discussion we misuse $\lambda$ for the
meaning of $|\lambda|$.
$N_{\rm cut}$ is 50 unless otherwise stated.

\subsection{Eigenvalue distribution}
\label{sec:eigenvalue_distribution}
\begin{figure}[tbp]
  \centering
  \includegraphics[width=10cm]{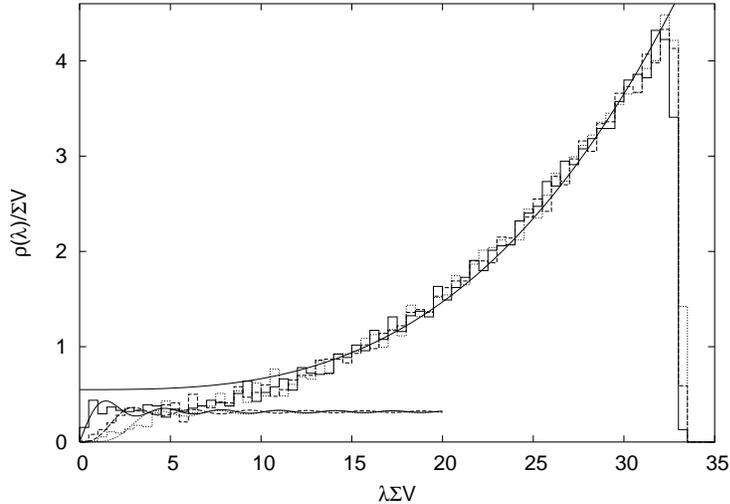}
  \caption{
    Eigenvalue distribution $\rho(\lambda)/\Sigma$ of the
    overlap-Dirac operator as a function of $\lambda\Sigma V$.
    Distributions are shown for topological sectors $|\nu|$
    = 0 (solid), 1 (dashed) and 2 (dotted).
    The asymptotic behavior $\sim\lambda^3$ in the bulk
    region is shown by a solid curve, while the fits with
    RMT are drawn by dashed curves.
  }
  \label{fig:distrib}
\end{figure}

First of all, we show the distribution of the
low-lying eigenvalues in Figure~\ref{fig:distrib}.
The plot shows the eigenvalue density
$\rho(\lambda) = 
 \left\langle \frac{1}{V}
 \sum_n \delta(\bar \lambda_n-\lambda)
 \right\rangle$
normalized by the chiral condensate $\Sigma$.
This normalization is determined for each topological sector
by comparing an average of the lowest eigenvalue with its
RMT expectation, as discussed below.

As expected from the dimensional analysis, the spectral
density behaves as $\sim\lambda^3$ in the large
$\lambda\Sigma V$ region.
This behavior is independent of the gauge field topology,
and a fit with $c_0+c_3(\lambda \Sigma V)^3$ 
for the entire gauge field
ensemble is shown by a solid curve
($c_0=0.55$, $c_3=1.15 \times 10^{-4}$).
Below $\lambda\Sigma V\simeq 10$, on the other hand, 
one can see a difference among different topological
sectors.
The curves show a fit with the expected functional
form from the chiral RMT.

\begin{figure}[tbp]
  $N_f=0$\\
  \scalebox{0.425}{\includegraphics{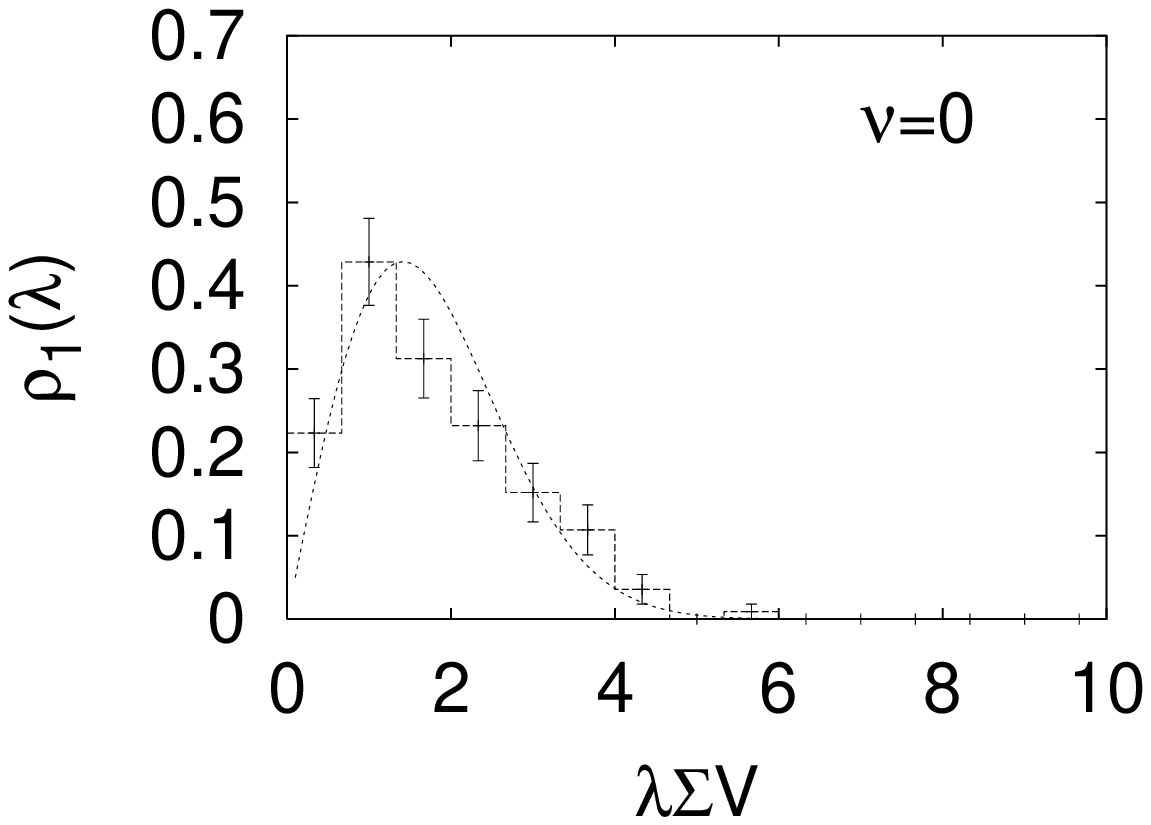}}
  \scalebox{0.425}{\includegraphics{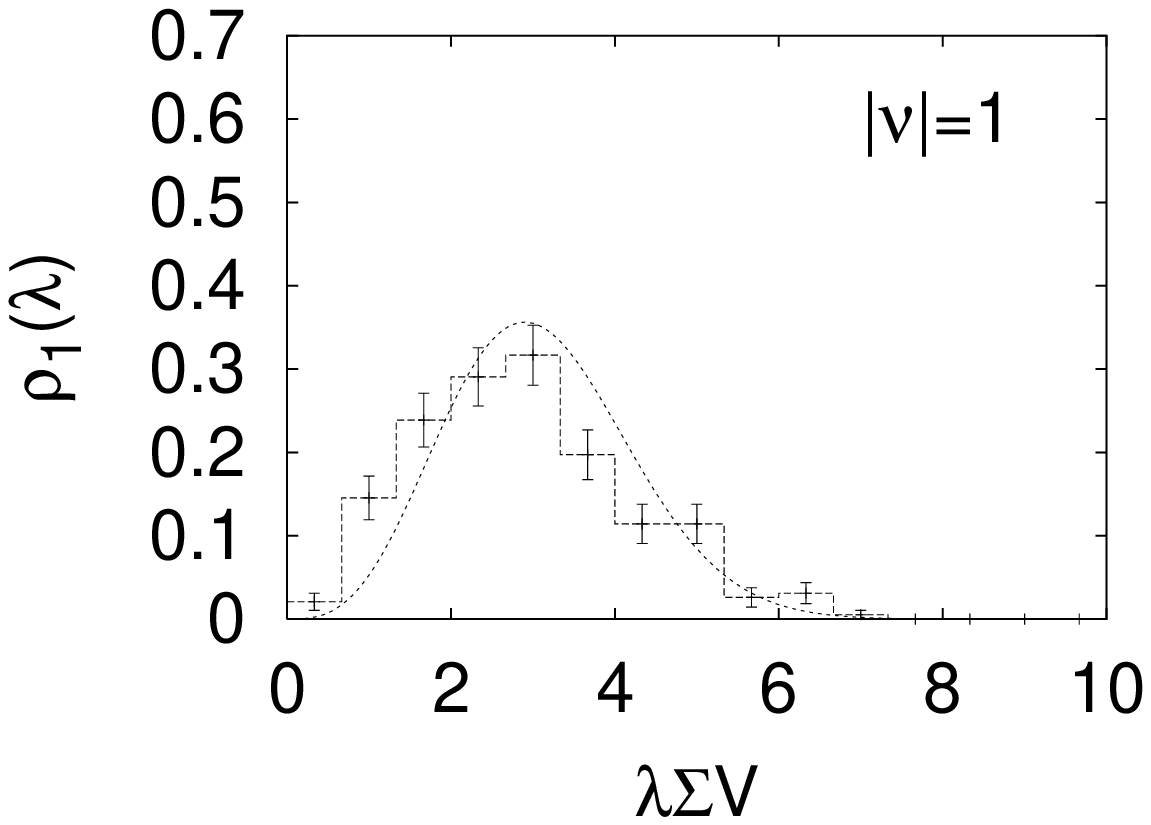}}
  \scalebox{0.425}{\includegraphics{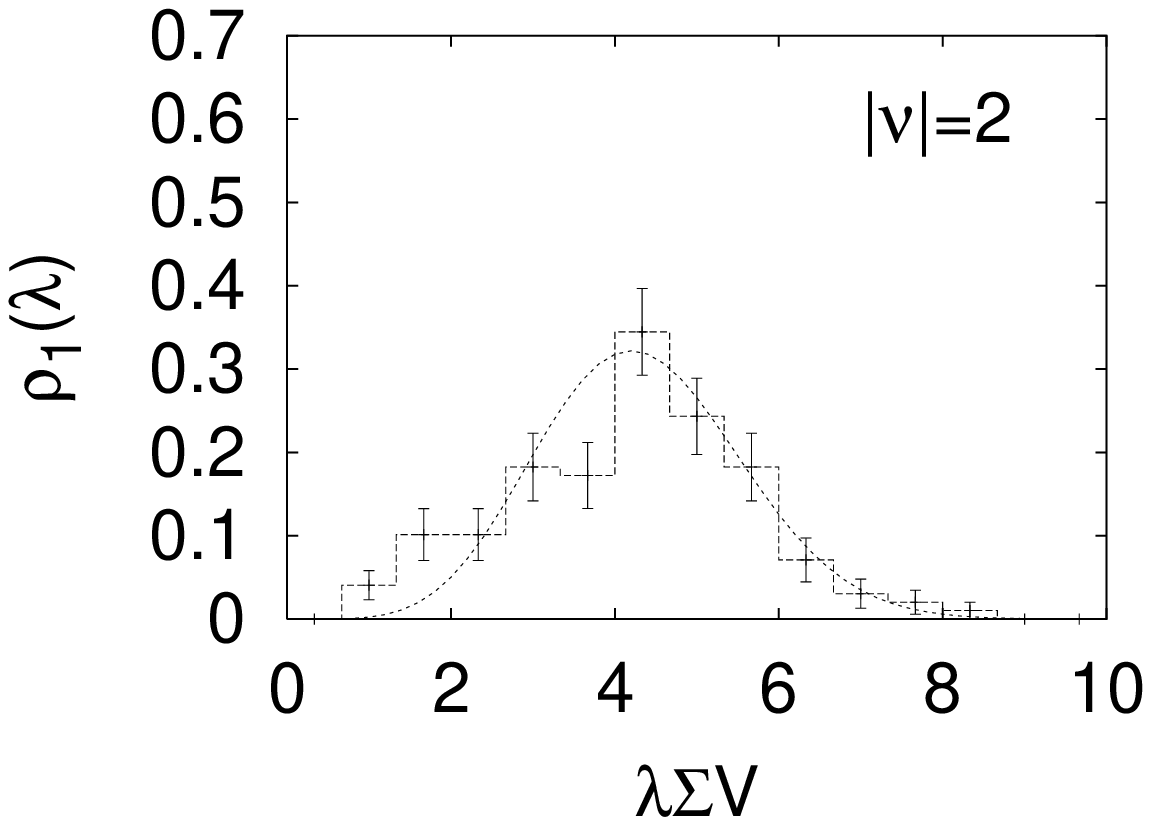}}
  \\
  $N_f=1$\\
  \scalebox{0.425}{\includegraphics{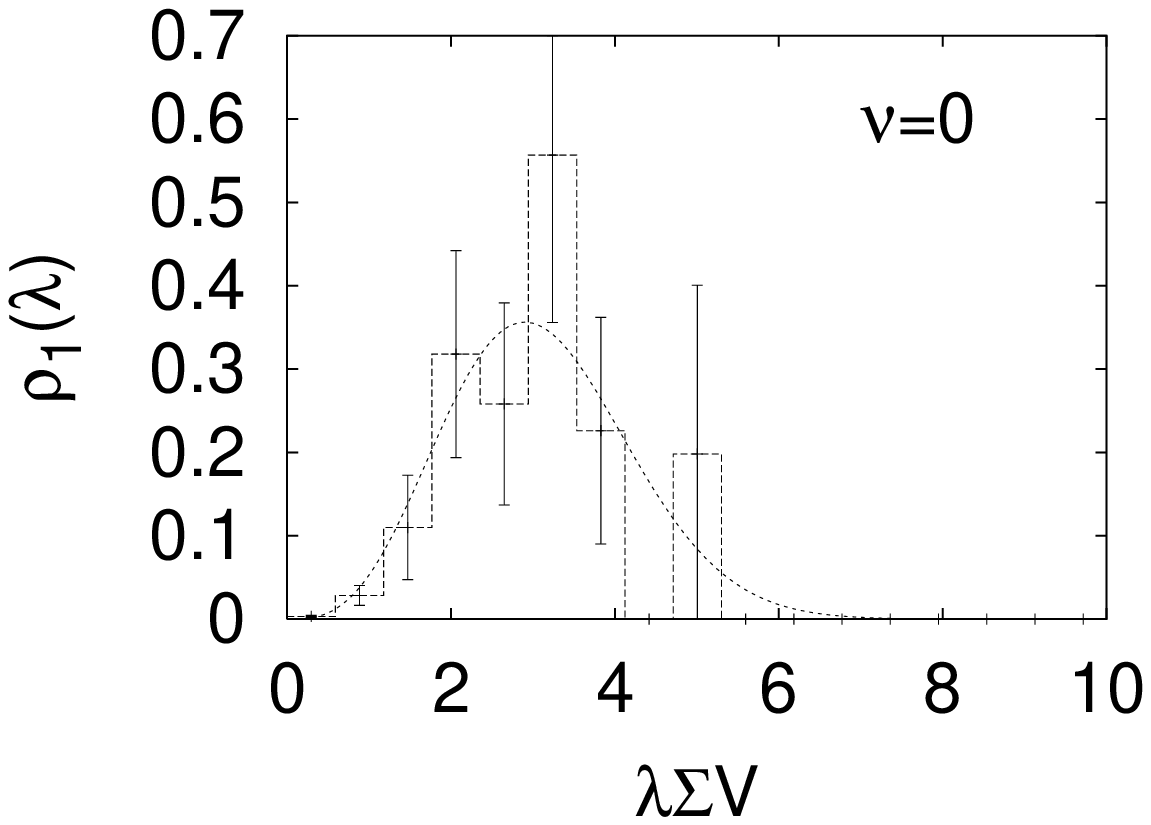}}
  \scalebox{0.425}{\includegraphics{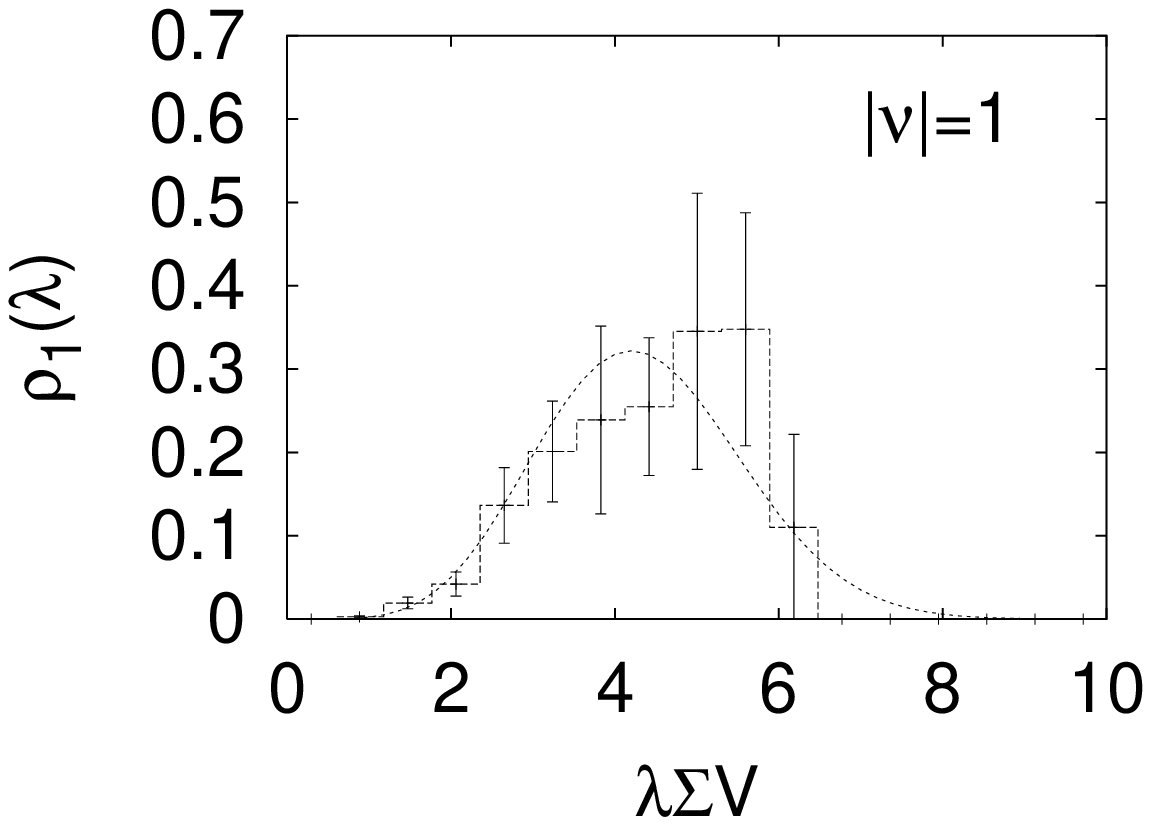}}
  \scalebox{0.425}{\includegraphics{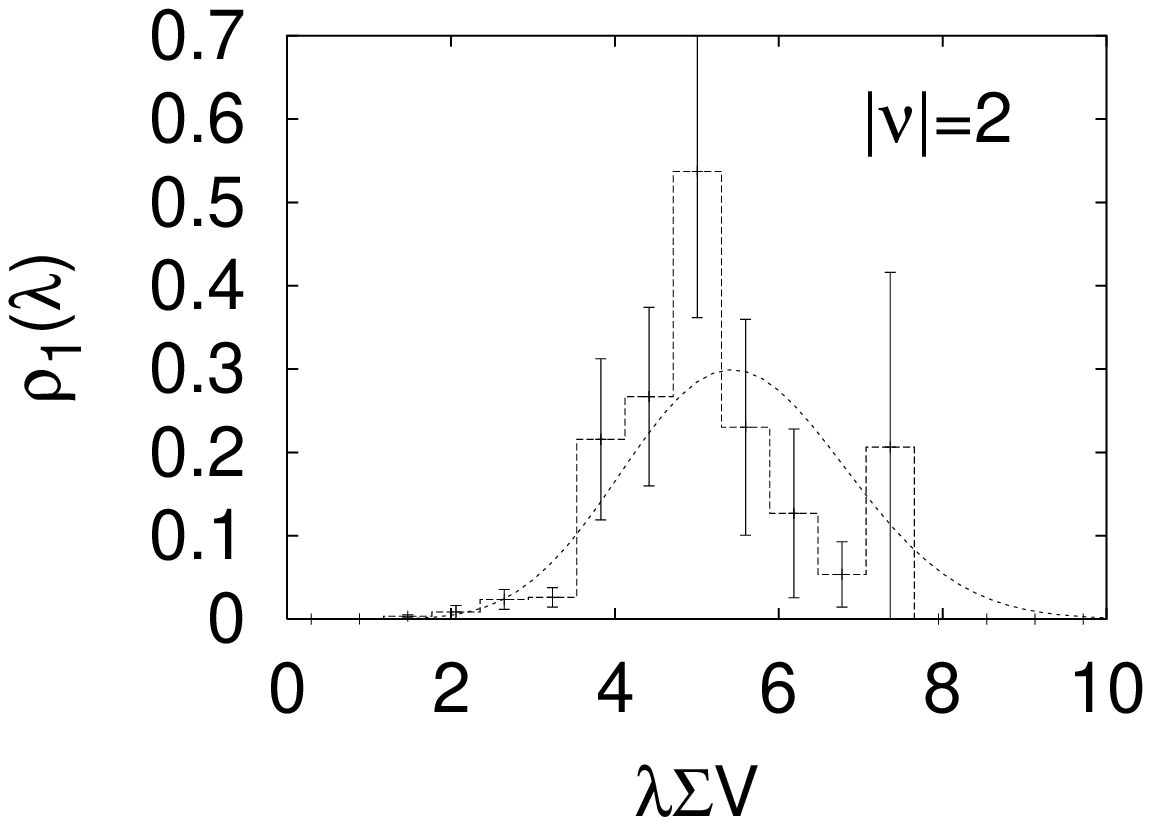}}
  \\
  $N_f=2$\\
  \scalebox{0.425}{\includegraphics{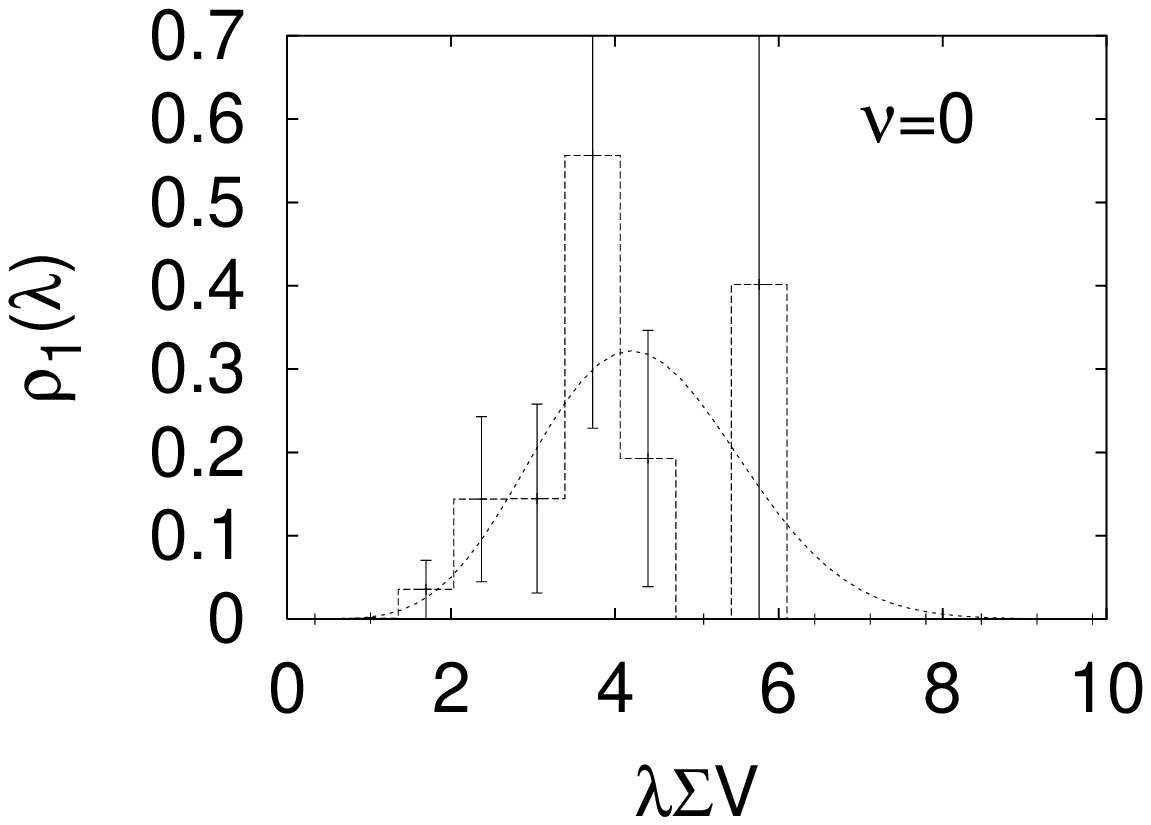}}
  \scalebox{0.425}{\includegraphics{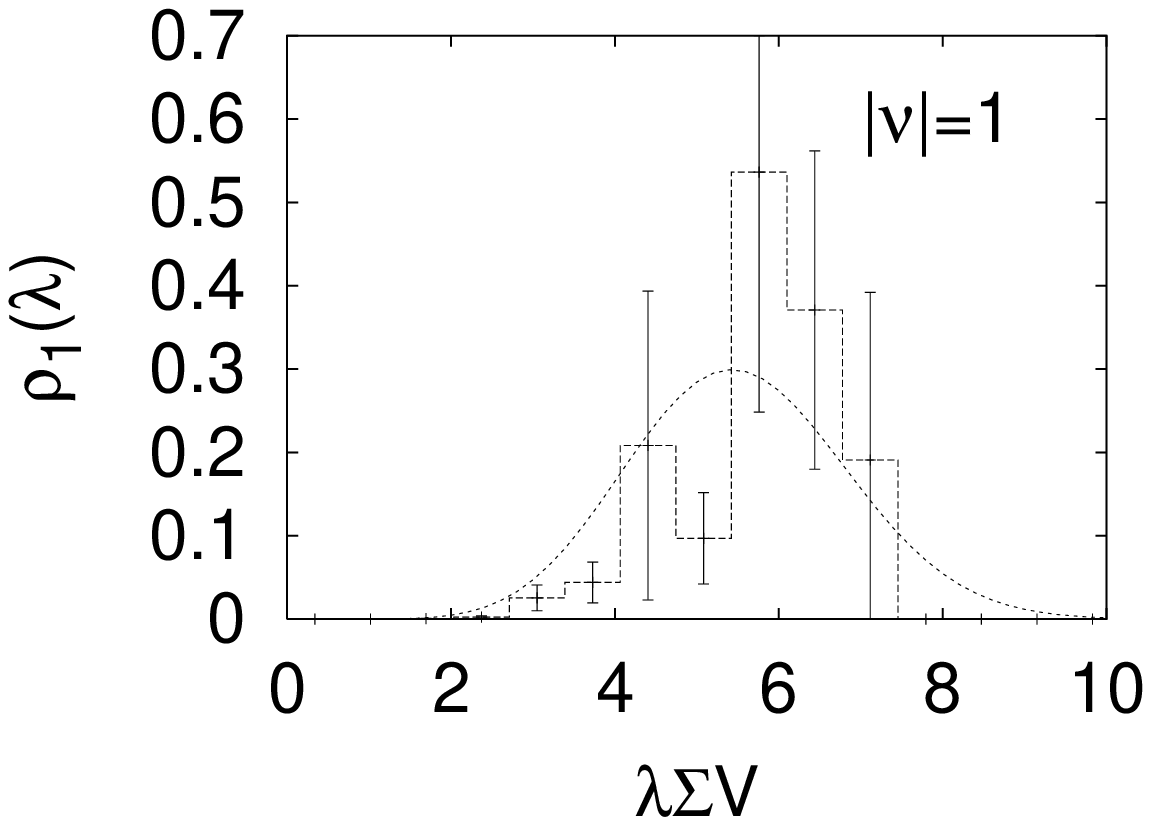}}
  \scalebox{0.425}{\includegraphics{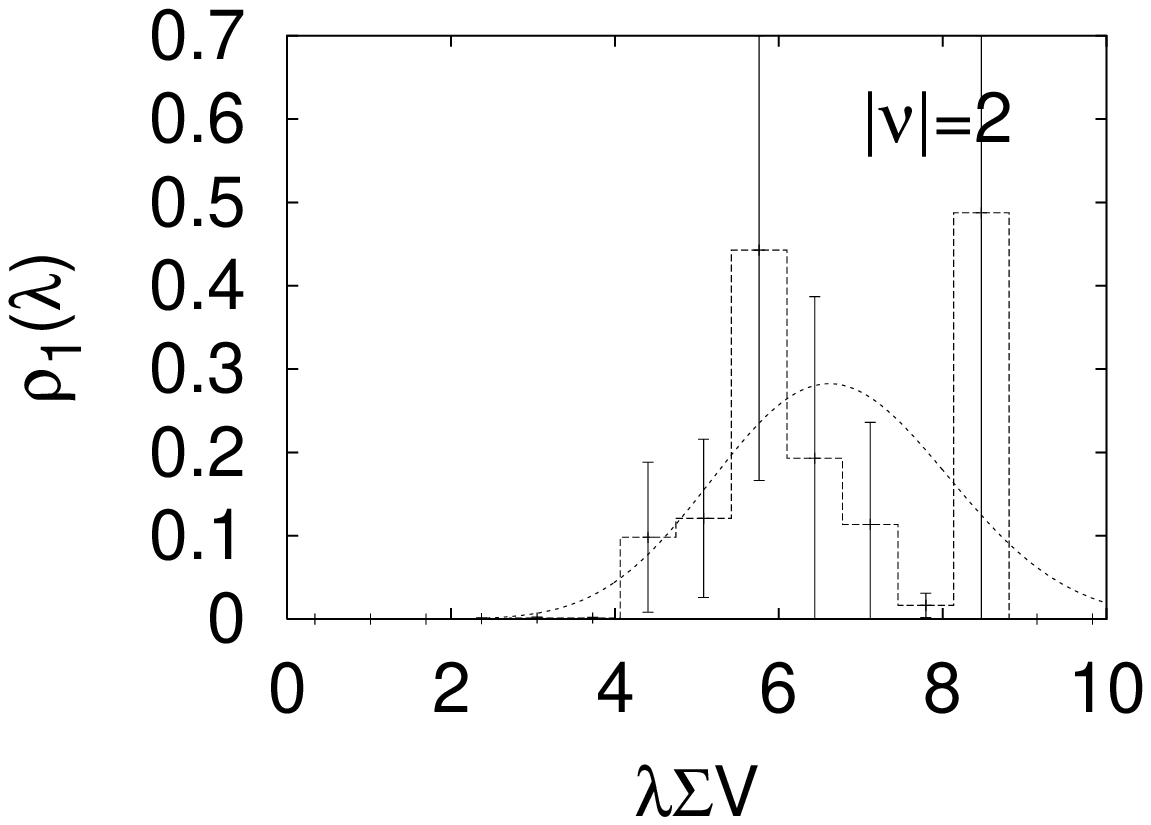}}
  \caption{
    Lowest non-zero eigenvalue distribution 
    on gauge configurations with a topological charge $\nu$ =
    0 (left), 1 (middle) and 2 (right).
    The results are shown for $N_f$ = 0 (top), 1 (middle), and
    2 (bottom). 
    The horizontal axis is normalized such that the average
    agrees with the RMT expectation.
  }
  \label{fig:lwstegn}
\end{figure}

In Figure~\ref{fig:lwstegn} we compare the distribution of
the lowest non-zero eigenvalue with the prediction from RMT
\cite{Nishigaki:1998is}.
Unlike the previous works
\cite{Edwards:1999ra,Bietenholz:2003mi,Giusti:2003gf},
we also obtain the distributions for non-zero number of
flavors, $N_f$ = 1 and 2.
For the quenched case, our results confirm the previous
works that found an agreement with the RMT predictions.

The unquenched distributions are obtained by the reweighting
with the truncated determinant.
The lowest eigenvalue distribution is probably one of the
most difficult quantities to be measured using the
reweighting method.
For instance, for $\nu=0$ almost all eigenvalues lie in the
region $\lambda\Sigma V\lesssim 5$ on the quenched lattice,
while only about a half of eigenvalues are expected to fall in
that region for $N_f$ = 2 as shown in the plot by a dotted
curve. 
The other half in the region $\lambda\Sigma V\gtrsim 5$ is
not sampled well with the reweighting.

In Table~\ref{table:lwtegn} we summarize the average of the 
lowest non-zero eigenvalue $\lambda_1$ predicted by RMT and
our results. 
For the RMT an expectation value of $\lambda_1\Sigma V$ is
listed, 
while the lattice data are bare values $\bar{a}\lambda_1$.
From these results we can extract the value of chiral
condensate $\Sigma$ for each $|\nu|$ and $N_f$.
For the lattice spacing we use the quenched value 
$a$ = 0.123~fm even for unquenched data, assuming that the
heavy quark potential is not much affected by the truncated
determinant.
The chiral condensate $\Sigma$ obtained in this work
corresponds to the bare lattice operator $\bar{\psi}\psi$
and is not renormalized.
To convert it to the continuum definition, such as the
$\overline{\mbox{MS}}$ scheme, we need a matching factor.

The agreement of $\Sigma$ given in Table~\ref{table:lwtegn}
among different topological sectors is encouraging.
The values with non-zero flavors are also in reasonable
agreement, though they are not necessarily agree since the
underlying theory is different.

\begin{table}[tbp]
\begin{center}
\begin{tabular}{|c|c|c|c|c|}\hline
  & $|\nu|$ & RMT ($\lambda_1\Sigma V$) 
  & Lattice ($\bar{a}\lambda_1$) & $\Sigma^{1/3}$ [MeV]
  \\
  \hline\hline
    &    0 & 1.77 & 0.0289(13) & 250(4)\\\cline{2-5}
$N_f$=0& 1 & 3.11 & 0.0486(12) & 254(2)\\\cline{2-5}
    &    2 & 4.34 & 0.0701(20) & 251(2)\\\hline \hline
    &    0 & 3.11 & 0.0569(59) & 241(8)\\\cline{2-5}
$N_f$=1& 1 & 4.34 & 0.0833(34) & 238(3)\\\cline{2-5}
    &    2 & 5.53 & 0.100(6)   & 242(5)\\\hline \hline
    &    0 & 4.34 & 0.067(12)  & 254(15)\\\cline{2-5}
$N_f$=2& 1 & 5.53 & 0.0939(42) & 247(04)\\\cline{2-5}
    &    2 & 6.69 & 0.112(14)  & 248(10)\\\hline
\end{tabular}
\caption{
  Expectation value of the lowest non-zero eigenvalue
  predicted by RMT $\lambda_1\Sigma V$ 
  and the lattice results $\bar a \lambda_1$.
  The chiral condensate $\Sigma$ is obtained with an input
  for lattice spacing $a$ = 0.123~fm at $\beta$ = 5.85.
}
\label{table:lwtegn}
\end{center}
\end{table}

\subsection{Partition function}
\label{sec:partition_function}
Employing the truncated determinant approximation, we
calculate the unquenched partition function at a fixed
topology $Z_\nu$ (\ref{eq:Znu_QCD}) as a function of the
quark mass.  

The number of eigenvalues included in the truncated
determinant has to be equal for different topological
charges in order to make the mass dimension consistent for
different topological sectors.
This is relevant for the partition function, because we
consider $Z_\nu/Z$, and $Z$ in the denominator is a sum
over all topological charges.
Therefore, we slightly modify the truncated determinant as 
$m^{|\nu|}
 \prod_{n=1}^{N_{\mathrm{cut}}-|\nu|/2}
 (\bar{\lambda}_n^2+m^2)$
when $|\nu|$ is even, or as
$m^{|\nu|}
 \prod_{n=1}^{N_{\mathrm{cut}}-(|\nu|+1)/2}
 (\bar{\lambda}_n^2+m^2)
 \times
 \sqrt{\bar{\lambda}_{N_{\mathrm{cut}}-(|\nu|-1)/2}^2 + m^2}$
when $|\nu|$ is odd.

\begin{figure}[tbp]
  \centering
  \includegraphics[width=10cm]{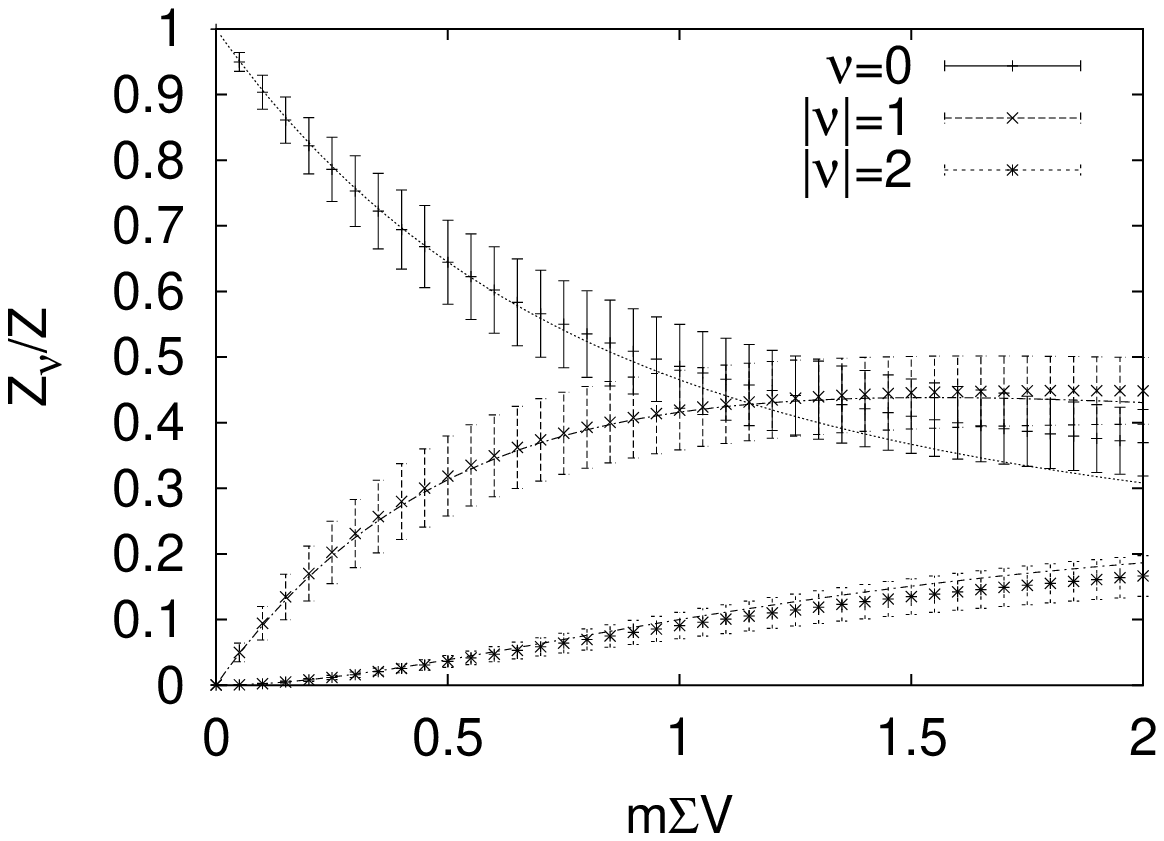}
  \includegraphics[width=10cm]{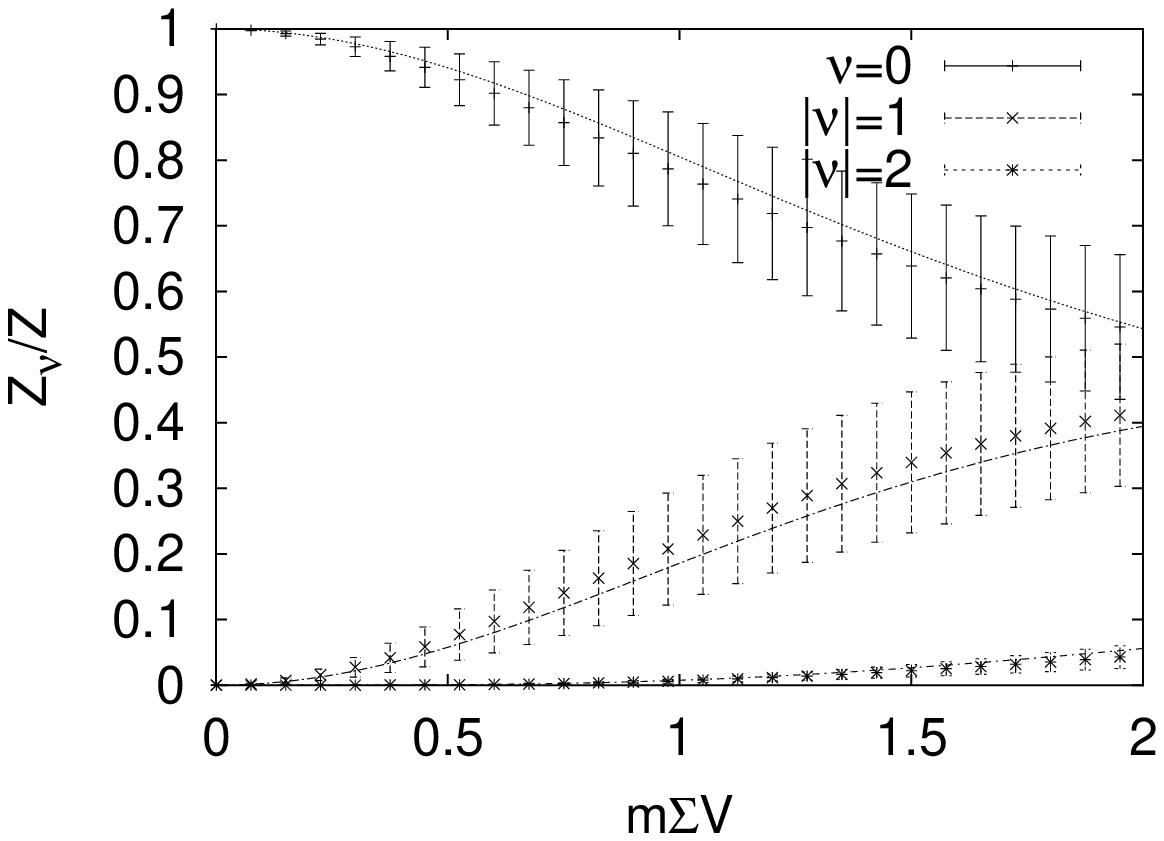}
  \caption{
    Partition function for a given topological charge as a
    function of the quark mass.
    The results are shown for $N_f$ = 1 (top) and 2 (bottom).
    The curves represent the analytic prediction in the
    $\epsilon$-regime. 
  }
  \label{fig:detmass_d8}
\end{figure}

In Figure~\ref{fig:detmass_d8} we plot $Z_\nu/Z$ as a
function of $m\Sigma V$ for $N_f$ = 1 (top panel) and 2
(bottom panel).
We find a good agreement with the expectation from the
chiral Lagrangian in the small quark mass region
(\ref{eq:Z_nu_Nf=1}) and (\ref{eq:Z_nu_Nf=2}).
A fit in the range
$\bar{a}m \leq$ 0.03 with a free parameter $\Sigma$
yields $\Sigma^{1/3}$ = 214(6)~MeV and 248(6)~MeV 
for $N_f$ = 1 and 2, respectively.
These results are quite insensitive to the value of 
$N_{\mathrm{cut}}$. 
They vary from 212(4)~MeV to 214(6)~MeV
by changing $N_{\rm cut}$ from 10 to 50 for $N_f$ = 1.
The variation is from 234(7)~MeV to 248(6)~MeV for $N_f$ =
2. 

\subsection{Eigenvalue sum rules}
Here we investigate the sum rules 
(\ref{eq:LS_sum-rule_1})-(\ref{eq:LS_sum-rule_2c}) 
of the Dirac operator eigenvalues.

\begin{figure}[tbp]
  sum rule I\hfill
  \raisebox{-0.5\height}{
    \includegraphics[scale=0.48]{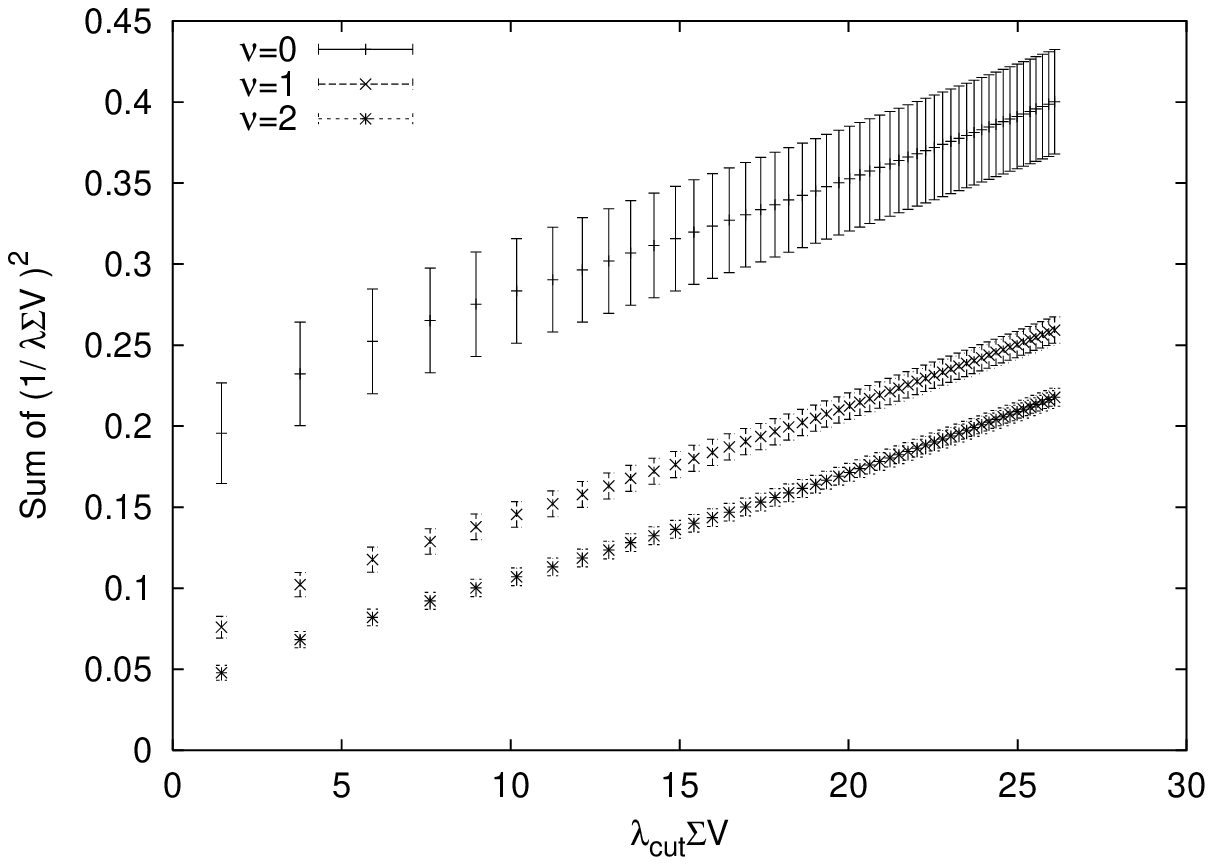}
    \includegraphics[scale=0.52]{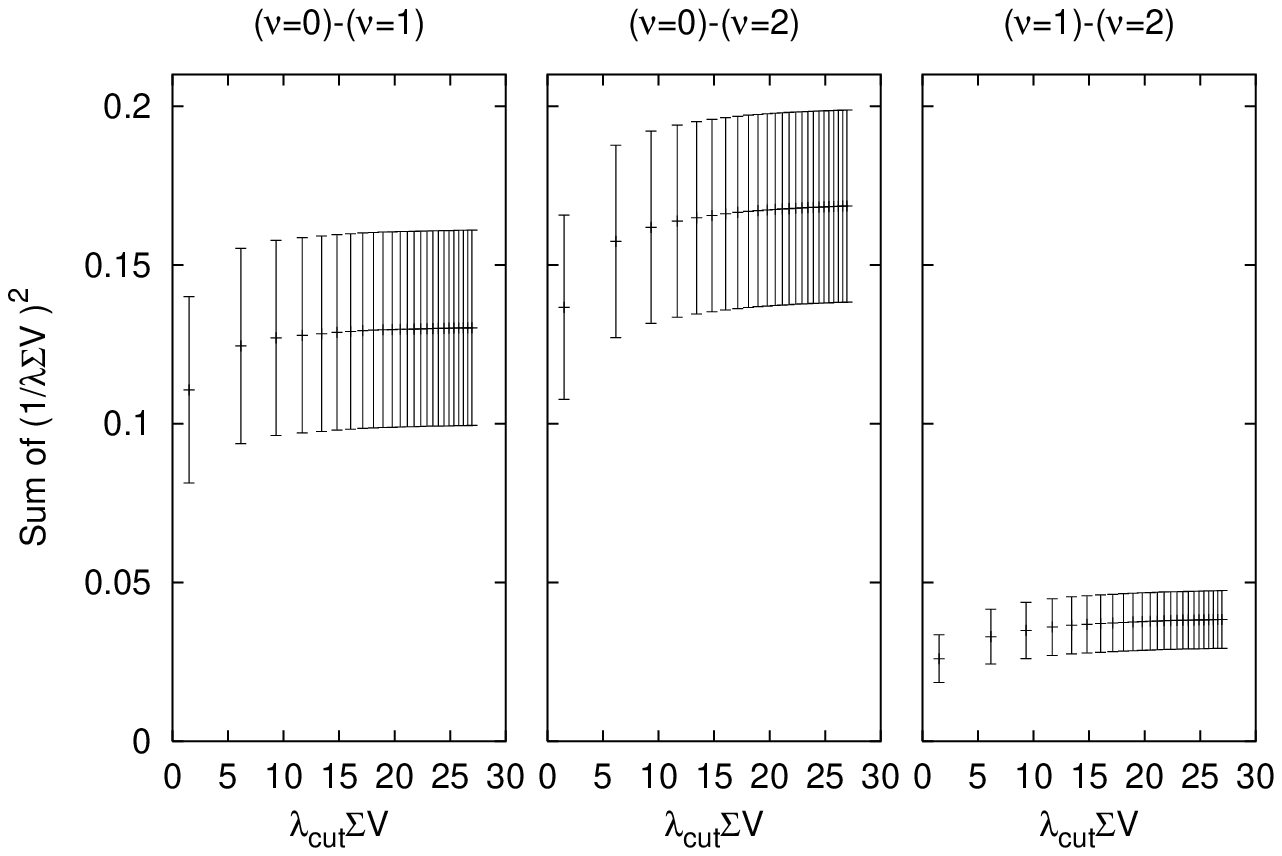}
  }\\[2mm]
  sum rule IIb\hfill
  \raisebox{-0.5\height}{
    \includegraphics[scale=0.48]{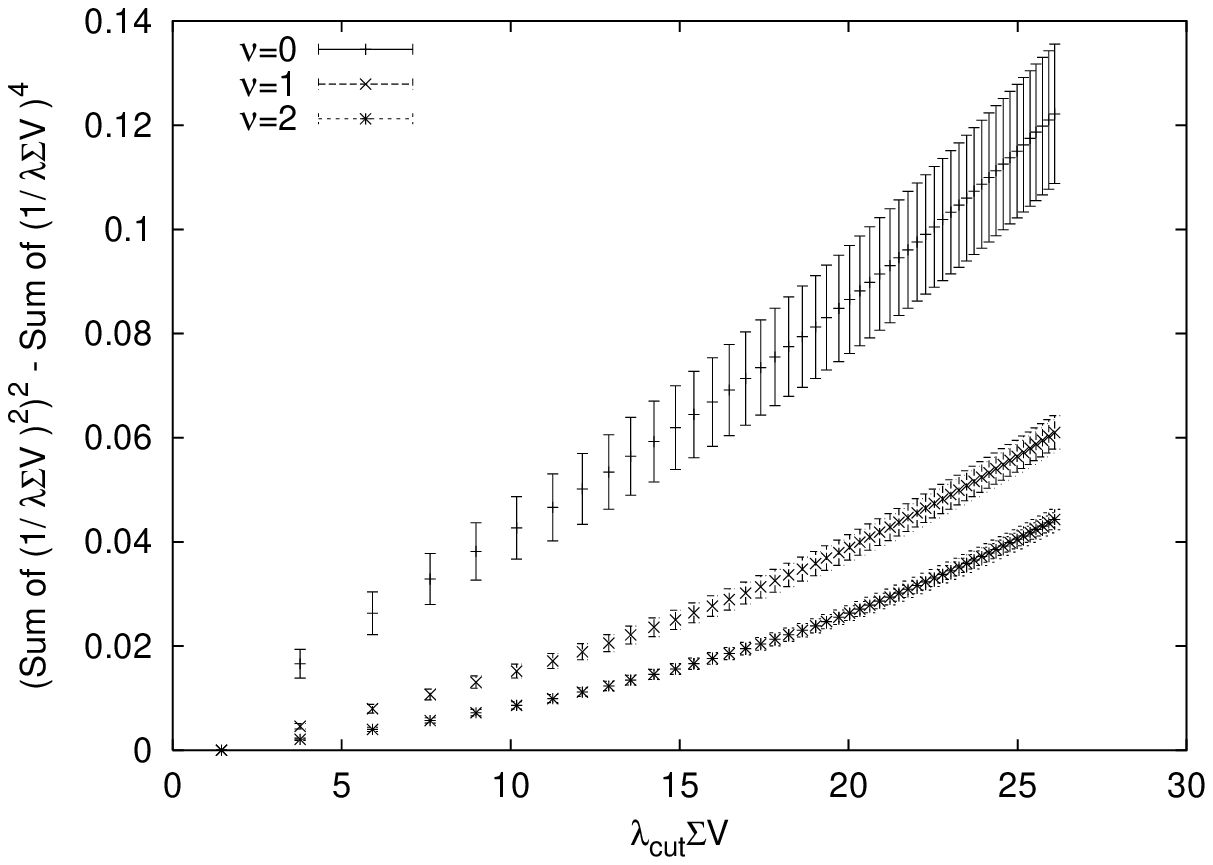}
    \includegraphics[scale=0.52]{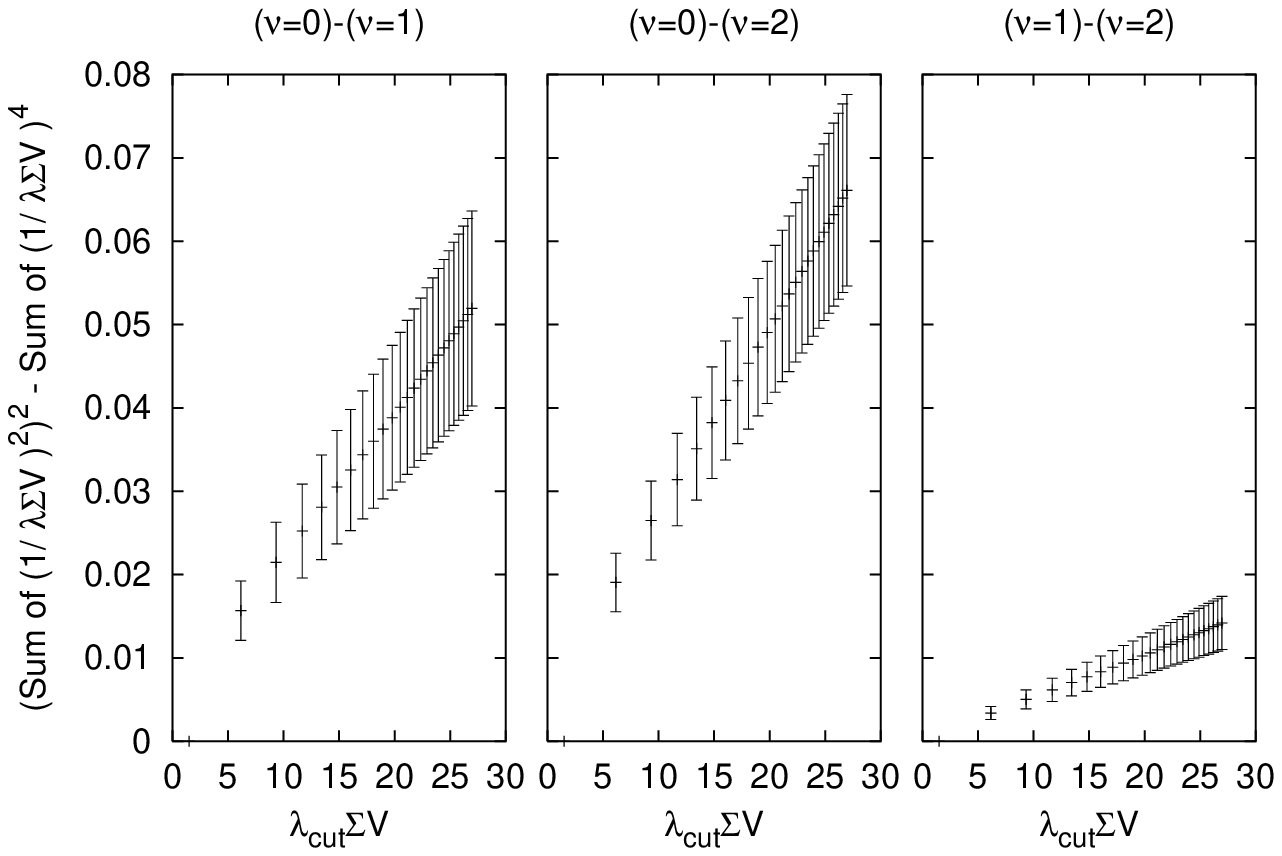}
  }\\
  \caption{
    Sum rules I (\ref{eq:LS_sum-rule_1}) and 
    IIb (\ref{eq:LS_sum-rule_2b}) for $N_f$ = 1.
    The ultraviolet divergence is shown on the left panels as
    a function of $\lambda_{\mathrm{cut}}\Sigma V$ with
    $\lambda_{\mathrm{cut}}$ the largest eigenvalue included
    in the sum. 
    The right panels show a saturation for the difference
    between different topological sectors.
  }
  \label{fig:sumrule_nf1}
\end{figure}

As discussed in Section~\ref{sec:Leutwyler-Smilga} the sums
$\sum_n^\prime 1/\lambda_n^2$ and $\sum_n^\prime 1/\lambda_n^4$
are divergent in the ultraviolet regime.
This behavior is explicitly shown in
Figure~\ref{fig:sumrule_nf1} for $N_f$ = 1.
The left panels are divergent sums as a function of
$\lambda_{\mathrm{cut}}\Sigma V$ with
$\lambda_{\mathrm{cut}}$ the largest eigenvalue included 
in the sum.
The sum rule I (top left) does not seem to show the
expected quadratic divergence but looks more like a linear
divergence. 
But, it is in fact consistent
with the integral
$\int^{\lambda_{\mathrm{cut}}} d\lambda
 \rho(\lambda)/\lambda^2$
with the form 
$\rho(\lambda)=\rho_\nu^{(0)}(\lambda)+\rho^{(3)}\lambda^3$ 
suggested in
Section~\ref{sec:eigenvalue_distribution}.
Here, $\rho^{(3)}\lambda^3$ represents the bulk distribution of
eigenvalues, which is independent of the gauge field
topology, while the $\rho_\nu^{(0)}(\lambda)$ term shows the
oscillating behavior depending on the topological charge.
For large $\lambda\Sigma V$, $\rho_\nu^{(0)}(\lambda)$
approaches to a constant $\rho^{(0)}$, which is
independent of the topology.
In the integral 
$\int^{\lambda_{\mathrm{cut}}} d\lambda
 \rho(\lambda)/\lambda^2$,
we observe that the curvature of 
$-\rho^{(0)}/\lambda_{\mathrm{cut}}$
cancels that of $\rho^{(3)} \lambda_{\mathrm{cut}}^2$ in the
region plotted in Figure~\ref{fig:sumrule_nf1}.

\begin{figure}[tbp]
  sum rule I\hfill
  \raisebox{-0.5\height}{
    \includegraphics[scale=0.48]{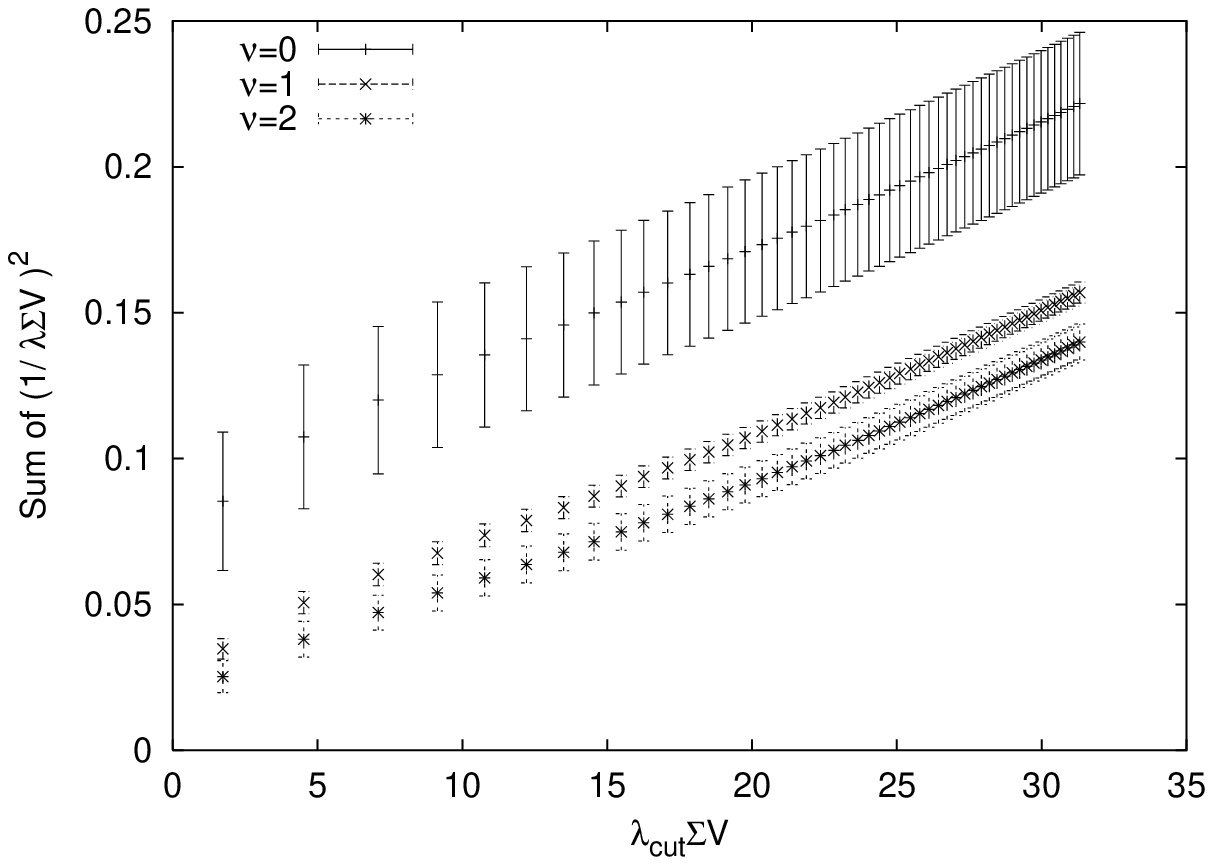}
    \includegraphics[scale=0.52]{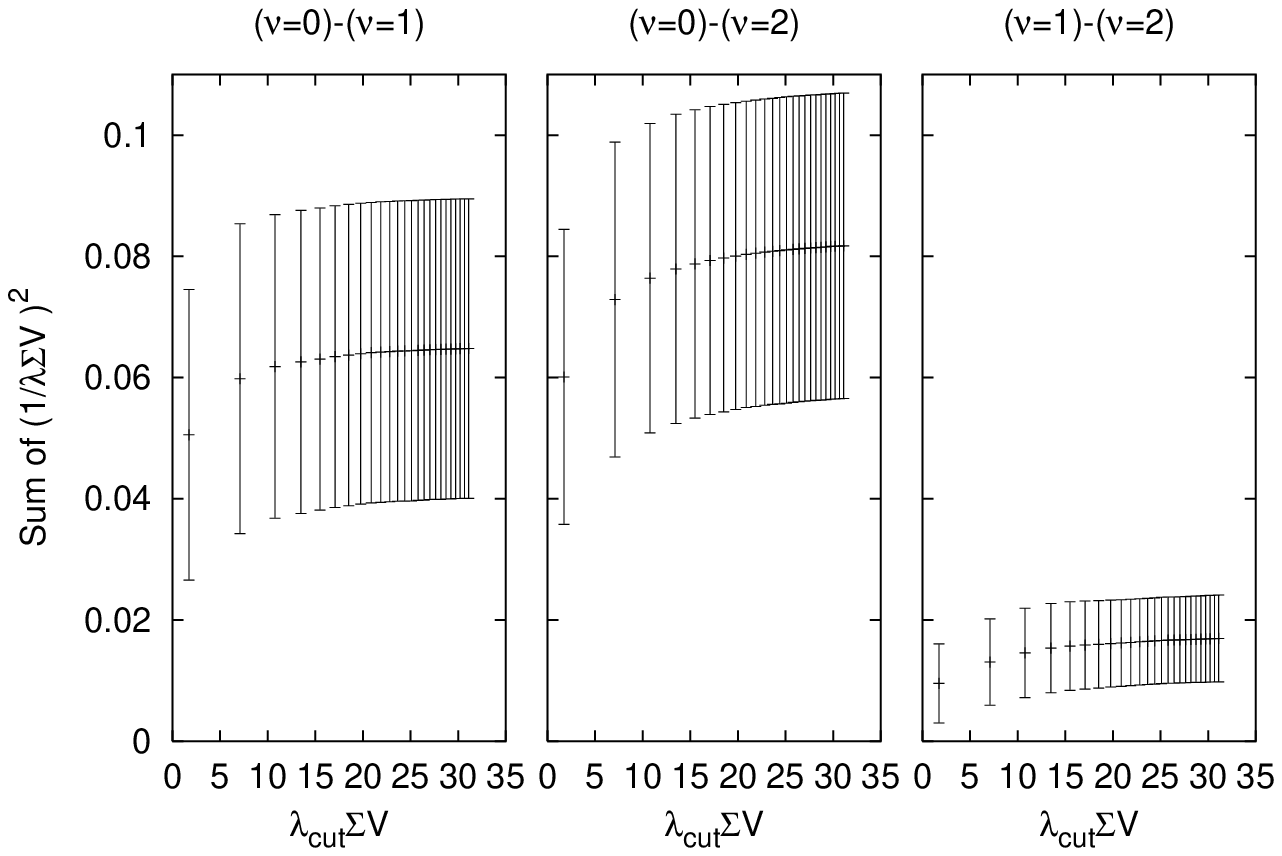}
  }\\[2mm]
  sum rule IIa\hfill
  \raisebox{-0.5\height}{
    \includegraphics[scale=0.48]{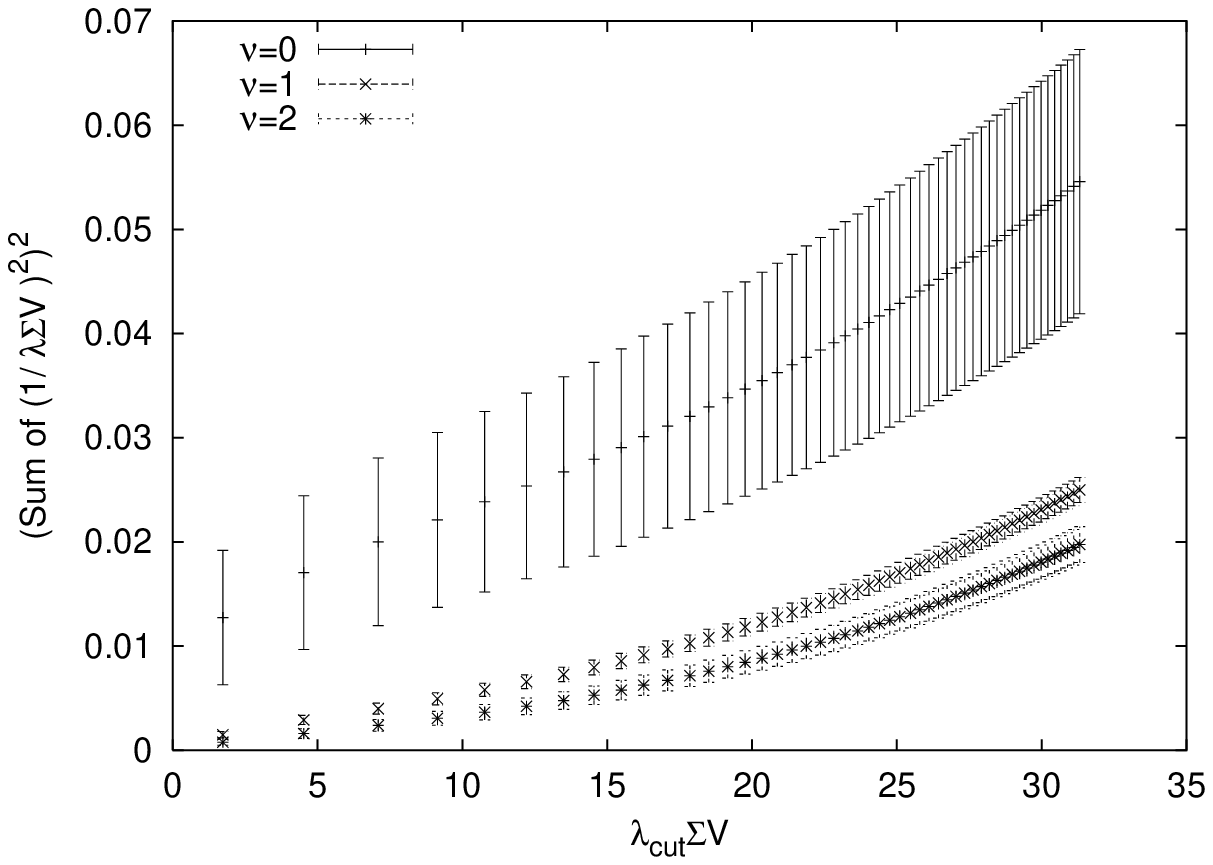}
    \includegraphics[scale=0.52]{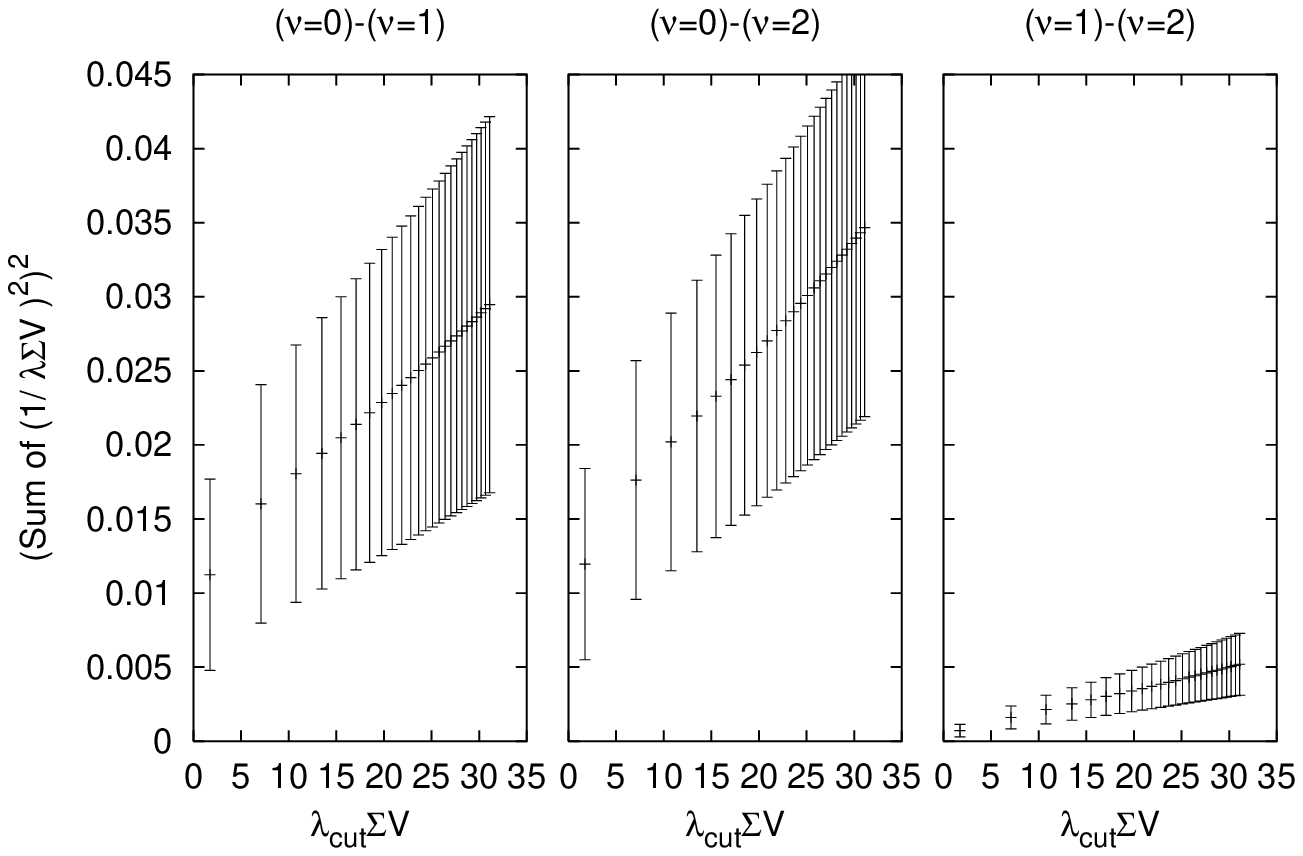}
  }\\[2mm]
  sum rule IIb\hfill
  \raisebox{-0.5\height}{
    \includegraphics[scale=0.48]{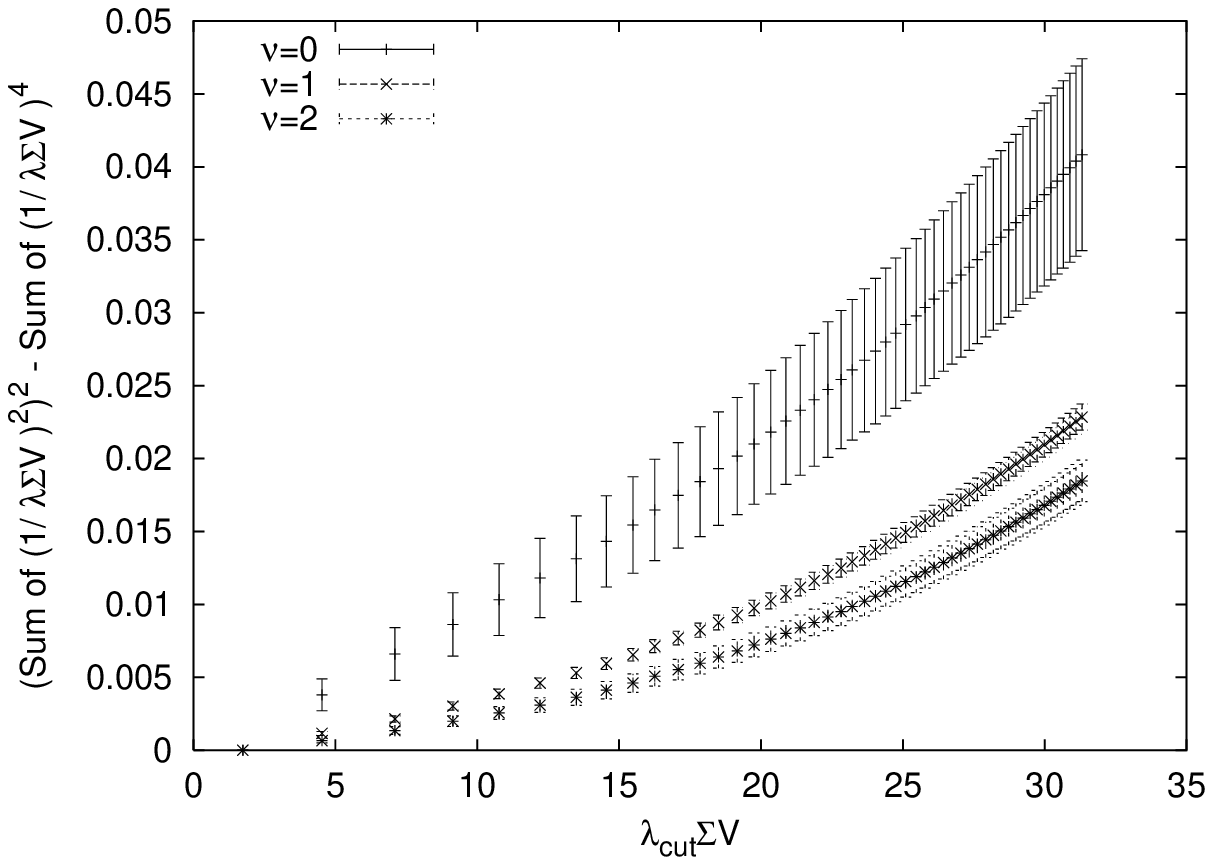}
    \includegraphics[scale=0.52]{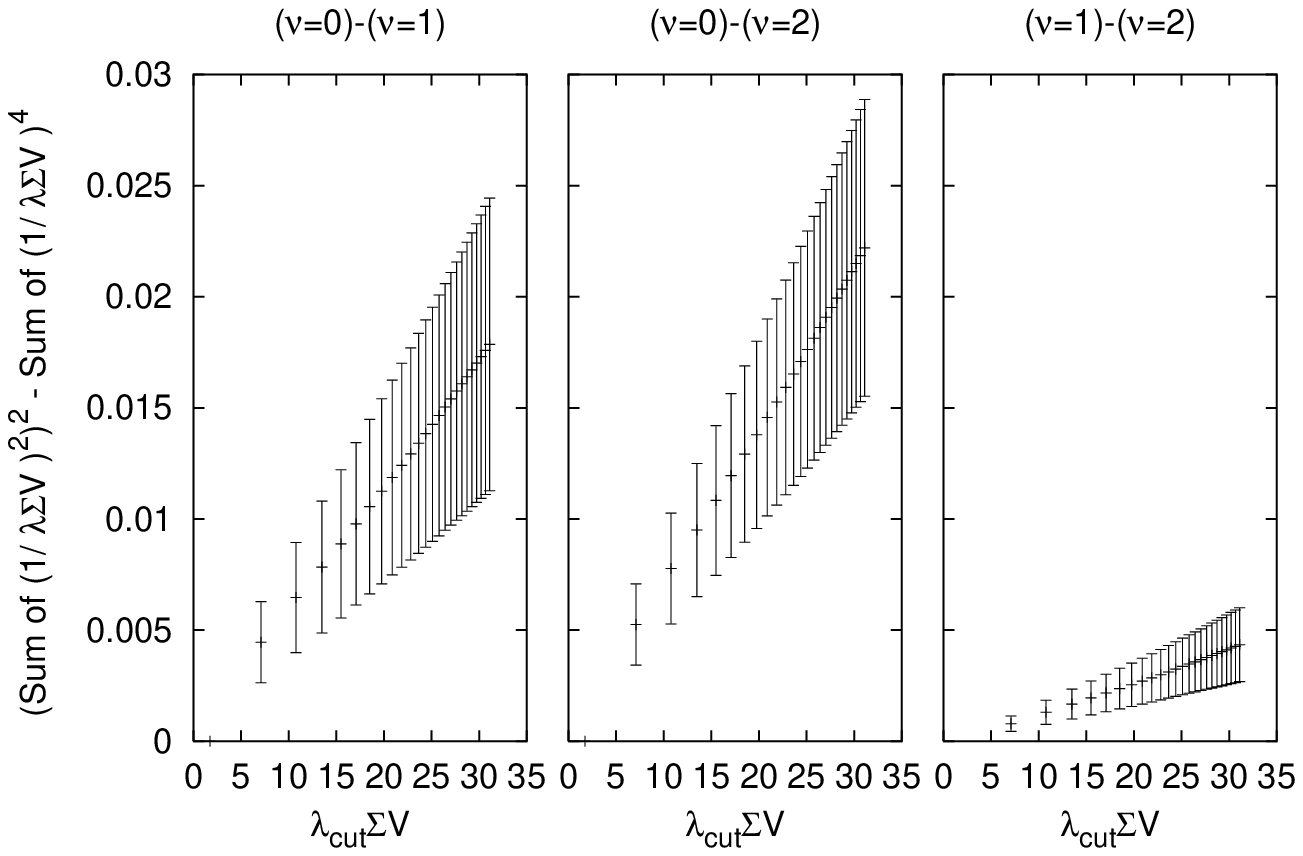}
  }\\[2mm]
  sum rule IIc\hfill
  \raisebox{-0.5\height}{
    \includegraphics[scale=0.48]{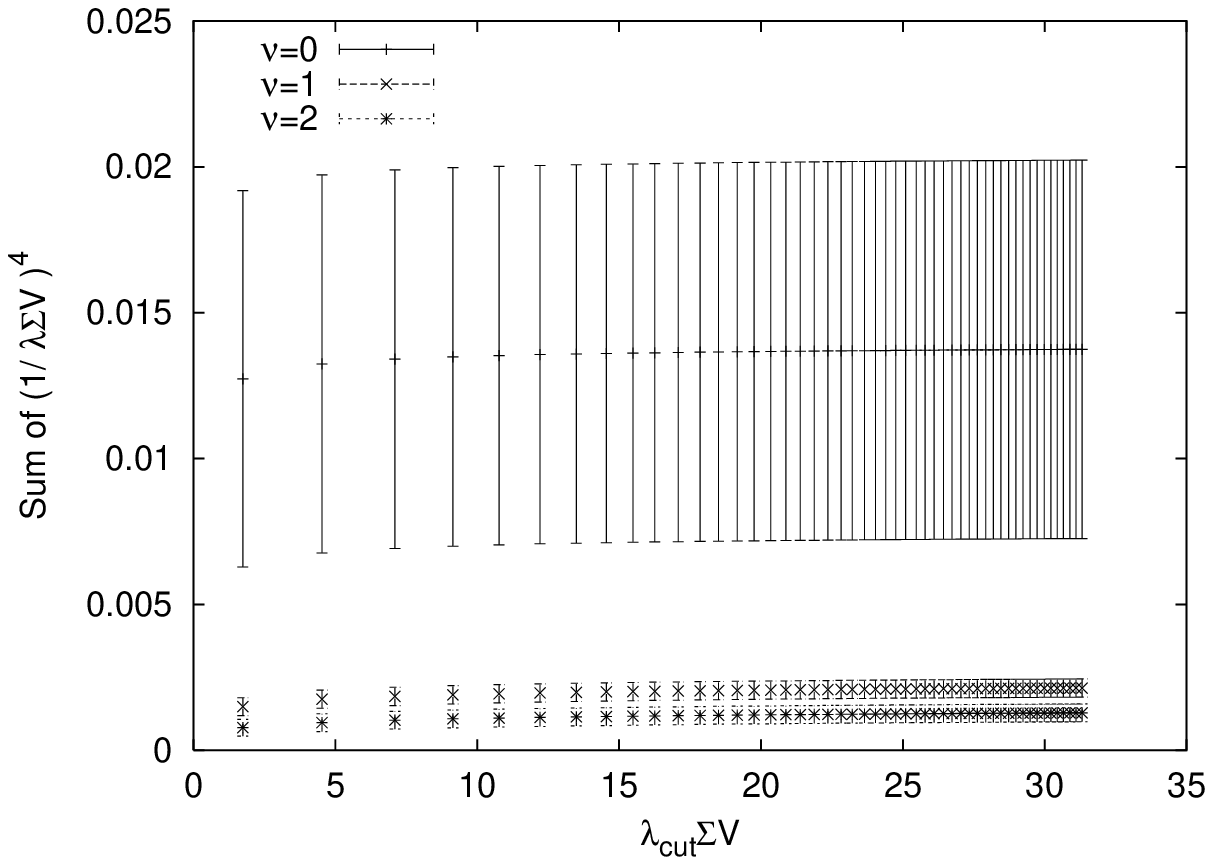}
    \includegraphics[scale=0.52]{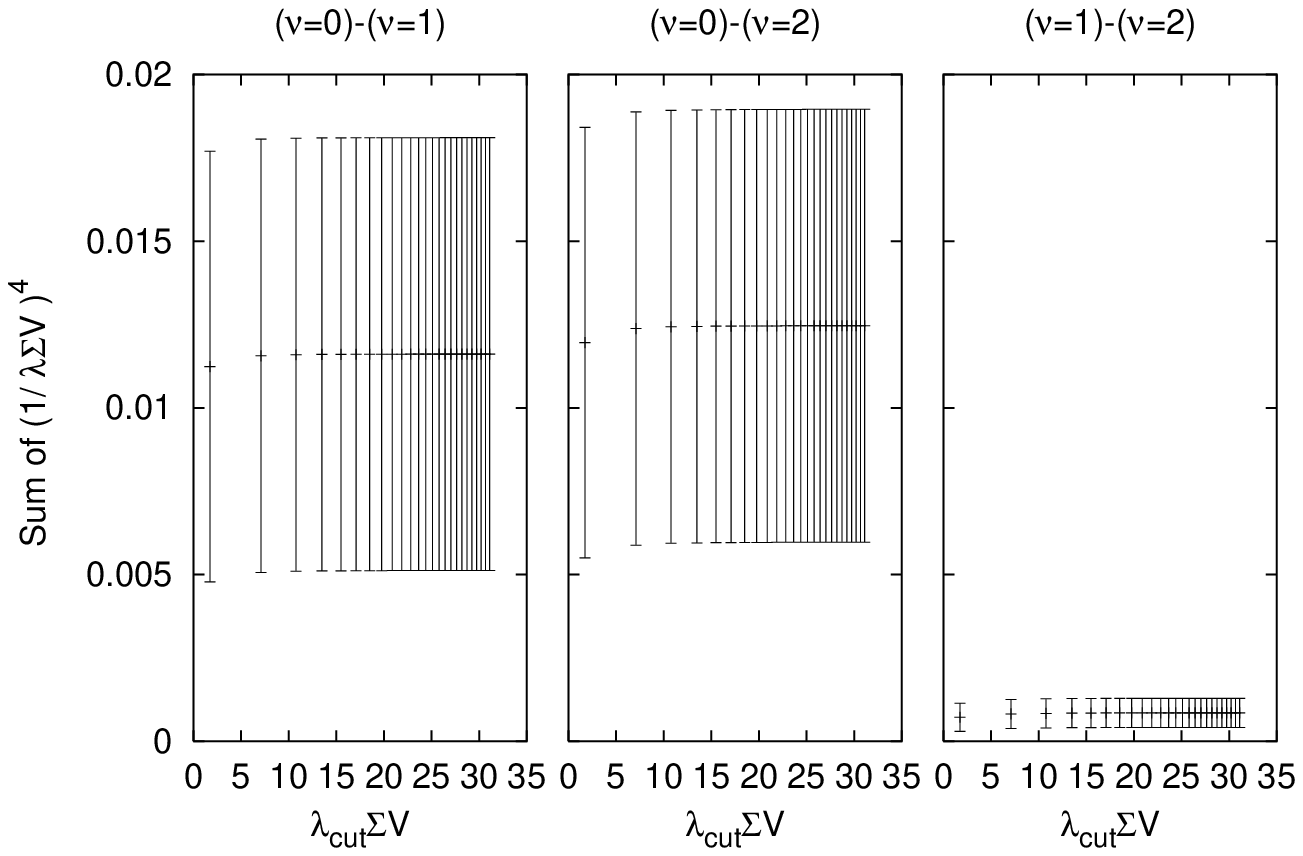}
  }\\
  \caption{
    Sum rules I (\ref{eq:LS_sum-rule_1}),
    IIa (\ref{eq:LS_sum-rule_2a}),
    IIb (\ref{eq:LS_sum-rule_2b}) and
    IIc (\ref{eq:LS_sum-rule_2c})
    for $N_f$ = 2.
    The ultraviolet divergence is shown on the left panels as
    a function of $\lambda_{\mathrm{cut}}\Sigma V$ with
    $\lambda_{\mathrm{cut}}$ the largest eigenvalue included
    in the sum. 
    The right panels show a saturation for the difference
    between different topological sectors.
  }
  \label{fig:sumrule_nf2}
\end{figure}

Similar plots are shown for $N_f$ = 2 in
Figure~\ref{fig:sumrule_nf2}.
The first order (sum rule I) and the second order (sum rules
IIa, IIb and IIc) are plotted.
The sum rule I is quadratically divergent as in the $N_f$ =
1 case; 
the divergence of the sum rule IIc (\ref{eq:LS_sum-rule_2c})
is mild as it is only logarithmic.
For the other second order sum rules (IIa and IIb), the
divergence is quartic, since it has the form 
$(\int^{\lambda_{\mathrm{cut}}} d\lambda
  \rho(\lambda)/\lambda^2)^2$.

In order to subtract these ultraviolet divergences, we
consider the differences of the eigenvalue sum rules among
different topological sectors.
This is motivated by an expectation that the bulk
distribution $ \rho^{(3)}\lambda^3$ is independent of the gauge
field topology as we can see in Figure~\ref{fig:distrib}.
On the other hand, the term $\rho_\nu^{(0)}$ carries the information of
the topology and shows the oscillating behavior expected
from RMT.
For the sum rule I (\ref{eq:LS_sum-rule_1}), to be explicit, 
the divergent piece are common for all topological sectors
and cancel in the differences.
In fact, the lattice data for the differences 
$(\nu=0)-(\nu=1)$, $(\nu=0)-(\nu=2)$ and $(\nu=1)-(\nu=2)$
are almost independent of the cutoff
$\lambda_{\mathrm{cut}}$ as shown in
Figure~\ref{fig:sumrule_nf1} (top right panel).
The same is true for the sum rule IIc as shown in
Figure~\ref{fig:sumrule_nf2} (4th row, right).

The cancellation of the ultraviolet divergence is not
expected for the sum rules IIa (\ref{eq:LS_sum-rule_2a}) and
IIb (\ref{eq:LS_sum-rule_2b}).
This is because they include a square of the integral
$(\int^{\lambda_{\mathrm{cut}}} d\lambda
  \rho(\lambda)/\lambda^2)^2$
and it contains a cross term
$\rho_\nu^{(0)}(\lambda) \times 
 \rho^{(3)} \lambda_{\mathrm{cut}}^2$, 
which is divergent but still topology-dependent through
$\rho_\nu^{(0)}(\lambda)$. 
The lattice data support this expectation as shown in 
Figure~\ref{fig:sumrule_nf1} (bottom right panel) and in 
Figure~\ref{fig:sumrule_nf2} (2nd and 3rd rows, right).

We use the sum rules I and IIc to extract $\Sigma$ from the
comparison of lattice data with the analytic formulae.
For $N_f$ = 1, only the sum rule I can be used, and we
obtain $\Sigma^{1/3}$ as
239(9), 238(7) and 233(9)~MeV
for the differences 
$(\nu=0)-(\nu=1)$, $(\nu=0)-(\nu=2)$ and $(\nu=1)-(\nu=2)$,
respectively. 
Similarly, for $N_f$ = 2 we obtain
268(17), 260(13) and 241(17)~MeV from the sum rule I, and 
257(12), 254(11) and 237(10)~MeV from the sum rule IIc.

\subsection{Topological susceptibility}
\begin{figure}[tbp]
\begin{center}
  \includegraphics[width=10cm]{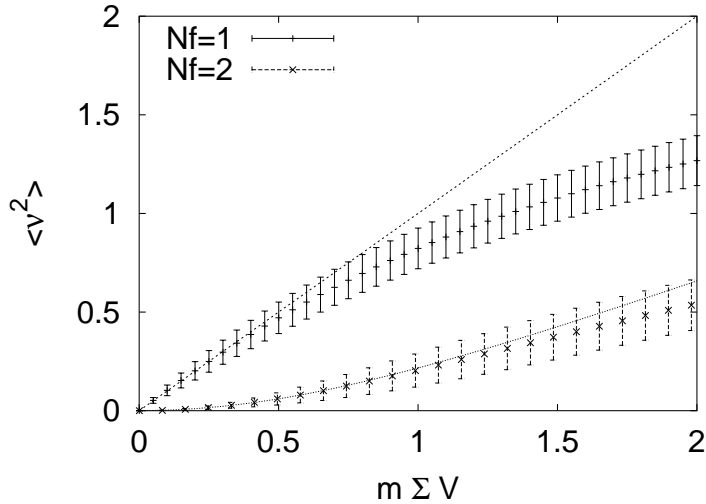}
  \caption{
    Expectation value of the topological charge squared
    $\langle\nu^2\rangle$ as a function of $m\Sigma V$.
    Results for $N_f$ = 1 (pluses) and 2 (crosses) are shown.
    The dashed line and the dotted curve are fit results with
    (\ref{eq:tpsus_nf1}) and (\ref{eq:tpsus_nf2}),
    respectively. 
    \label{fig:tpsus}
  }
\end{center}
\end{figure}

The mass dependence of the topological susceptibility
$\langle\nu^2\rangle/V$ in the $\epsilon$-regime is
analytically known as 
(\ref{eq:tpsus_nf1}) and (\ref{eq:tpsus_nf2}).
Lattice calculation of this quantity is straight-forward
from the partition function discussed in
Section~\ref{sec:partition_function}.
In Figure~\ref{fig:tpsus} we plot $\langle\nu^2\rangle$
as a function of $m\Sigma V$.
The value of $\Sigma$ extracted by fitting in the region
$\bar{a}m\leq 0.03$ is 
$\Sigma^{1/3}$ = 216(9)~MeV for $N_f$ = 1, and 257(18)~MeV
for $N_f$ = 2. 
These results are quite stable under the change of
$N_{\mathrm{cut}}$.
The variation for $N_{\mathrm{cut}}$ = 10--50 is
215(7)--216(9)~MeV for $N_f$ = 1 and
246(27)--257(18)~MeV for $N_f$ = 2.

The clear mass dependence of the topological susceptibility
as observed in Figure~\ref{fig:tpsus} has not been obtained
before in the dynamical fermion simulations
\cite{Durr:2001ty}, except an exploratory work by 
Kov\'acs \cite{Kovacs:2001bx}, which utilized the same
reweighting technique as employed in our work.
The reason is partly that the quark mass is not small enough
to realize the $\epsilon$-regime. 
A recent work with slightly smaller quark masses shows an
indication of the suppression of the topological
susceptibility \cite{Allton:2004qq}.
Also, the (improved) Wilson fermion formulation employed in
the previous dynamical simulations does not respect the
chiral symmetry, and it is questionable if the expected
quark mass dependence is obtained at finite lattice
spacings. 
For the improved staggered dynamical quarks, with which one
can reach the small quark mass region, there is an
indication that the expected quark mass dependence is
reproduced in the continuum limit \cite{Bernard:2003gq}

\subsection{Comparison}
In Table~\ref{table:chcnd} we summarize the values of chiral
condensate $\Sigma$ extracted from our numerical data.
We set $a$ = 0.123~fm from the Sommer scale $r_0$ at $\beta$
= 5.85.

First of all, the quenched ($N_f$ = 0) results obtained from
the lowest eigenvalue distribution are consistent among
different topological sectors, and in perfect agreement with
a previous work by Bietenholz {\it et al.}
\cite{Bietenholz:2003mi}, who used the same lattice
parameters, {\it i.e.} $\beta$ = 5.85 and 10$^4$ lattice,
and extracted $\Sigma$ from the lowest eigenvalue
distribution as $\Sigma^{1/3}\simeq$ 253~MeV.
In \cite{Fukaya:2005yg} we investigated the meson
correlators in the $\epsilon$-regime at the same $\beta$
value but on a larger lattice 10$^3\times$20.
The quenched chiral condensate is extracted from the scalar
and pseudo-scalar meson correlators as $\Sigma^{1/3}\simeq$
257(14)~MeV, which is also consistent with the present work.

For $N_f$ = 1, we find an agreement between the results
from the lowest eigenvalue distribution and those from the
sum rule I.
The results are around $\Sigma^{1/3}\simeq$ 240~MeV, which
is in the same ballpark with the quenched results.
On the other hand, the fits for the partition function and
the topological susceptibility yield substantially lower
values, 214(6)~MeV and 216(9)~MeV, respectively.
At first sight, this disagreement seems puzzling, because
the eigenvalue sum rules are derived through the derivatives
of the partition function with respect to $m^2$.
The main difference is that the partition function and the
topological susceptibility involve the relative weight among
different topological sectors.
Namely, the leading mass dependence of the partition
function $Z_\nu\propto m^{|\nu|}$, which is most relevant to
the relative weight, is factored out before the sum rules
are derived.
The eigenvalue sum rules represent the non-leading mass
dependence.
Therefore, the disagreement implies that the relative weight
among different topological sectors is no well described by
(\ref{eq:Z_nu_Nf=1}).

In contrast to the $N_f$ = 1 case, the results for
$\Sigma$ from several different methods nicely agree with
each other for $N_f$ = 2.
Probably, the extraction from the lowest eigenvalue
distribution contains substantial systematic error due to
the sampling problem, as discussed in
Section~\ref{sec:eigenvalue_distribution}.
The sum rules are less problematic, because they involve the
inverse of the low-lying eigenvalues and its most important
contribution comes from lower side of the distribution,
which is well sampled with the reweighting method.

\begin{table}[tbp]
\begin{center}
  \begin{tabular}{|c|c|c|}
    \multicolumn{3}{l}{$N_f$ = 0}\\
    \hline
                      &$|\nu|=0$ & 250(4) MeV \\ \cline{2-3}
    lowest eigenvalue &$|\nu|=1$ & 254(2) MeV \\ \cline{2-3}
                      &$|\nu|=2$ & 251(2) MeV \\ \hline
    \multicolumn{3}{l}{$N_f$ = 1}\\
    \hline
                      &$|\nu|=0$ & 241(8) MeV \\ \cline{2-3}
    lowest eigenvalue &$|\nu|=1$ & 238(3) MeV \\ \cline{2-3}
                      &$|\nu|=2$ & 242(5) MeV \\ \hline
               &$(|\nu|=0)-(|\nu|=1)$ & 239(9) MeV \\ \cline{2-3}
    sum rule I &$(|\nu|=0)-(|\nu|=2)$ & 238(7) MeV \\ \cline{2-3}
               &$(|\nu|=1)-(|\nu|=2)$ & 233(9) MeV \\ \hline     
    partition function         & all & 214(6) MeV \\ \hline
    topological susceptibility & all & 216(9) MeV \\ \hline
    \multicolumn{3}{l}{$N_f$ = 2}\\
    \hline
                      &$|\nu|=0$ & 254(15) MeV \\ \cline{2-3}
    lowest eigenvalue &$|\nu|=1$ & 247(4)  MeV \\ \cline{2-3}
                      &$|\nu|=2$ & 248(10) MeV \\ \hline     
               &$(|\nu|=0)-(|\nu|=1)$ & 268(17) MeV \\ \cline{2-3}
    sum rule I &$(|\nu|=0)-(|\nu|=2)$ & 260(13) MeV \\ \cline{2-3}
               &$(|\nu|=1)-(|\nu|=2)$ & 241(17) MeV \\ \hline     
                 &$(|\nu|=0)-(|\nu|=1)$ & 257(12) MeV \\ \cline{2-3}
    sum rule IIc &$(|\nu|=0)-(|\nu|=2)$ & 254(11) MeV \\ \cline{2-3}
                 &$(|\nu|=1)-(|\nu|=2)$ & 237(10) MeV \\ \hline     
    partition function         & all & 248(6)  MeV \\ \hline
    topological susceptibility & all & 257(18) MeV \\ \hline
  \end{tabular}
  \caption{
    Summary of the results for chiral condensate
    $\Sigma^{1/3}$.
    Results from the lowest eigenvalue distribution,
    eigenvalue sum rules I and IIc,
    partition function, and topological susceptibility
    are listed for $N_f$ = 0, 1 and 2.
    The second column indicates the topological charge of
    the gauge configuration used for the measurement.
  }
  \label{table:chcnd}
\end{center}
\end{table}

\section{Conclusions}
The effect of the fermion determinant to the QCD vacuum is
substantial for small enough quark masses.
In the language of chiral perturbation theory, the constant
modes dominate the low energy dynamics, when the system is
confined in a finite volume.
The quark mass dependence of the partition function can be
analytically derived using the $\epsilon$-expansion in the
chiral perturbation theory.
The analytic formulae by Leutwyler and Smilga show strong
dependence on the topological charge and the number of
dynamical quark flavors. 
In this numerical work we explicitly tested these analytic
expectations using lattice QCD simulations.

The fermion determinant is approximated by a truncated
product of low-lying eigenvalues of the overlap-Dirac
operator. 
The eigenvalues are truncated at around 1200~MeV.
Intuitively, this approximation should be effective as far
as the low energy dynamics is concerned, and in fact the
relative weight among gauge configurations is roughly
unchanged when we further increase the cutoff for the
eigenvalue.

From our lattice calculations of 
(i) lowest-lying eigenvalue distribution, 
(ii) eigenvalue sum rules,
(iii) quark mass dependence of the partition function, and
(iv) topological susceptibility,
we found good agreement with the analytic predictions.
For these quantities, the only free parameter in the chiral
perturbation theory (at the leading order) is the chiral
condensate $\Sigma$, which we can extract from the lattice
data.
The extraction of $\Sigma$ from the methods (i)--(iv)
provides a highly non-trivial cross-check of the theory (or
the lattice calculation), and our results at $N_f$ = 2
are all consistent with each other.
A problem remains at $N_f$ = 1, for which the extraction
with a fixed topological charge ((i) and (ii)) and the fits
to all topological charges ((iii) and (iv)) are mutually
inconsistent.
It may signal a breakdown of the $\epsilon$-expansion at
$N_f$ = 1, for which no Nambu-Goldstone massless pion exists
and therefore the finite momentum excitation is not large
compared to the ground state energy.

Although the reweighting method with the truncated
determinant provides a good qualitative approximation, 
its limitation is also apparent.
First of all, the systematic uncertainty due to the
truncation is hardly estimated.
For the quantities (i)--(iv) we explicitly checked that the
results are unchanged by varying the number of eigenvalues
from 10 to 50, but it does not guarantee that they remain
unchanged until the limit of the exact determinant is
reached.
This problem could be solved either by defining an effective
Dirac operator which includes the low-lying eigenvalues only
or by incorporating the effects of the rest of the
eigenmodes using stochastic methods.
Second, the Monte Carlo sampling with the pure gauge action
becomes ineffective when quark mass becomes small.
Near the massless limit, we loose nearly 95\% (98\%) of the
statistics for $N_f$ = 1 (2) by the suppression factor due
to the fermion determinant. 
This problem can only be cured by including the fermion
determinant into the Boltzmann weight when the gauge
configuration is generated.

\section*{Acknowledgements}
We thank Hidenori Fukaya for many useful discussions.


%



\end{document}